\documentclass[journal]{IEEEtran}
\usepackage[T1]{fontenc}

\usepackage{amssymb}
\usepackage{amsmath}
\usepackage{cite}
\usepackage{url}
\usepackage{cite,graphicx,amsmath,amssymb}
\usepackage{fancyhdr}
\usepackage{mdwmath}
\usepackage{mdwtab}
\usepackage{caption}
\usepackage{amsthm}
\usepackage{setspace}
\usepackage{hyperref}
\usepackage{algorithm}
\usepackage{algorithmic}
\usepackage{multirow}
\usepackage{makecell}
\usepackage{mathtools}
\usepackage{subcaption}
\usepackage{bm}
\usepackage{tikz}
\usepackage{xurl}
\usepackage{stfloats}
\usepackage{mathrsfs}

\usepackage{booktabs}
\usepackage{multirow}
\usepackage{siunitx}
\usepackage{balance}
\usepackage{bm}

\usepackage{array}
\usepackage{multirow,tabularx}
\usepackage{varwidth}

\hypersetup{colorlinks=true,
	linkcolor=blue,
	citecolor=blue,      
	urlcolor=black,
}

\expandafter\def\expandafter\normalsize\expandafter{%
	\normalsize%
	\setlength\abovedisplayskip{6pt}%
	\setlength\belowdisplayskip{6pt}%
	\setlength\abovedisplayshortskip{3.4pt}%
	\setlength\belowdisplayshortskip{3.4pt}%
}

\theoremstyle{definition}

\newtheorem{theorem}{Theorem}
\newtheorem{lemma}{Lemma}
\newtheorem{remark}{Remark}

\newtheorem{proposition}{Proposition}

\allowdisplaybreaks

\graphicspath{{./figure/}}

\begin{document}
	
	\title{Center-Fed Pinching Antenna System (C-PASS): Modeling, Analysis, and Beamforming Design}

	\author{Xu Gan,~\IEEEmembership{Member, IEEE}, and Yuanwei Liu,~\IEEEmembership{Fellow,~IEEE} 		
		\thanks{(Corresponding author: Yuanwei Liu.)}
		\thanks{Xu Gan is with the Department of Electrical and Computer Engineering, The University of Hong Kong, Hong Kong (e-mail: eee.ganxu@hku.hk).}
		\thanks{Yuanwei Liu is with the Department of Electrical and Computer Engineering, The University of Hong Kong, Hong Kong, and also with the Department of Electronic Engineering, Kyung Hee University, Yongin-si, Gyeonggi-do 17104, South Korea (e-mail: yuanwei@hku.hk).}
	}
	
	\maketitle
	
	\begin{abstract}
		A generalized framework for the novel center-fed pinching antenna system (C-PASS) is proposed. Within this framework, closed-form expressions for the degree of freedom (DoF) and power scaling law of the proposed C-PASS are first derived. These theoretical results reveal that the achievable DoF scales linearly with the number of input ports, $M$, and the number of receive antennas, $K$. Furthermore, the derived power scaling laws demonstrate that the C-PASS achieves a power gain of order $\mathcal{O}(P_T M)$, where $P_T$ denotes the transmit power. Based on the proposed C-PASS modeling, a sum-rate maximization problem for the joint optimization of transmit and pinching beamforming is then formulated. To solve this highly coupled non-convex problem, an efficient alternating optimization algorithm is developed. More particularly, the transmit precoding and power splitting ratios are updated via derived closed-form solutions, while the pinching antenna positions and radiation coefficients are optimized using block coordinate descent (BCD) methods. Finally, our numerical results reveal that the single-waveguide C-PASS: 1) achieves superior DoF and power scaling laws compared to the single-waveguide PASS; and 2) outperforms the multi-waveguide PASS in high-attenuation regimes, yielding a substantial gain exceeding $10$ dB.

		
	\end{abstract}

	\section{Introduction}
	\IEEEPARstart{T}{he} evolution towards beyond sixth-generation (B6G) networks anticipates unprecedented spectral efficiency and ubiquitous connectivity~\cite{saad2019vision,akyildiz20206g}. To fulfil these demands, the pinching-antenna system (PASS)~\cite{liu2025pinching,liu2025pinching2,liu2026survey} has emerged as a promising flexible-antenna paradigm. Specifically, PASS feeds radio frequency (RF) signals into dielectric waveguides for long-distance transmission, which are subsequently radiated by pinching antennas (PAs) deployed in the vicinity of users. In this setup, the signal propagation within the waveguide can be regarded as a stable line-of-sight (LoS) channel with negligible attenuation. Leveraging this characteristic, PASS effectively substitutes a significant portion of the high-loss free-space propagation with low-loss wired transmission, thereby substantially shortening the effective wireless distance~\cite{tyrovolas2025performance,xu2025pinching2}. Consequently, this architecture mitigates path loss and blockage probability, yielding a highly deterministic channel dominated by the LoS component\cite{ouyang2026capacity,xie2026pinching}.

	Nevertheless, the high reliability of the in-waveguide LoS channel comes at the cost of spatial multiplexing capability. More particularly, the strong LoS dominance inevitably results in a rank-one equivalent channel, thereby restricting the available degrees of freedom (DoF) to $\text{DoF}=1$. It is noteworthy that this DoF bottleneck is intrinsic to the conventional PASS architecture. Thus, the system DoF remains saturated at one and fails to scale with the number of deployed PAs or feeding signals\cite{yang2025pinching}. This restriction prevents the exploitation of spatial multiplexing gains, rendering the single-waveguide PASS unable to support multiple parallel data streams to serve multiple users or realize multi-functional capabilities. To be specific, this bottleneck imposes severe challenges across three critical functions: 1) For multi-user communications\cite{shan2025exploiting}, the rank deficiency results in inseparable signal superposition, preventing the recovery of individual user streams; 2) For channel acquisition\cite{11018390}, the high spatial correlation hinders the accurate separation and estimation of multipath channel state information (CSI); and 3) For wireless sensing\cite{wang2025wireless}, the limited spatial DoF restricts angular and ranging resolutions.

	\subsection{Prior Works}
	Motivated by these challenges, enhancing the spatial DoF of PASS-assisted communications constitutes a pivotal research direction for the practical deployment of future wireless networks. To address this, existing literature can be primarily classified into three categories: 1) multi-waveguide conventional PASS; 2) multi-waveguide novel PASS; and 3) single-waveguide center-fed PASS (C-PASS).

	\subsubsection{Multi-waveguide Conventional PASS}
		One straightforward strategy to expand the multiplexing capability is to employ multiple waveguides. In this configuration, the system DoF scales linearly with the number of deployed waveguides. Exploiting this scalability, the authors of \cite{wang2025modeling} utilized $N$ dielectric waveguides to serve $K$ users, where the condition $N \ge K$ ensures sufficient spatial DoF for simultaneous multi-user transmission. Under this framework, a joint transmit and pinching beamforming scheme was developed to minimize the total transmit power. Adopting a similar multi-waveguide configuration, the work in \cite{bereyhi2025mimo} shifted the focus to maximizing the downlink weighted sum-rate via joint precoding and PA placement, while further designing an iterative multi-user detection algorithm for uplink transmissions. To further enhance architectural flexibility, the authors of~\cite{zhao2025pinching} introduced three waveguide-connection strategies, namely waveguide multiplexing, waveguide division, and waveguide switching. Tailored for multi-group multicast systems, these architectures were optimized using a joint baseband and pinching beamforming design. Extending to multi-functional wireless networks, recent research has explored diverse applications of multi-waveguide PASS\cite{li2025mimo,papanikolaou2025physical,wang2025joint}. As for simultaneous wireless information and power transfer (SWIPT), the authors of~\cite{li2025mimo} leveraged multiple waveguides to simultaneously support information decoding and energy harvesting receivers. This study maximized the information sum-rate while guaranteeing the minimum harvested energy constraints by jointly optimizing pinching beamforming and PA positions. Investigating physical layer security, the work in \cite{papanikolaou2025physical} proposed an artificial noise-aided beamforming framework. By jointly designing information beams, noise covariance matrices, and PA coordinates, the secrecy rate against eavesdroppers was maximized. Furthermore, for symbiotic radio applications, the authors of~\cite{wang2025joint} formulated a joint optimization problem to maximize the achievable sum-rate, considering detection error probability constraints of backscatter devices and antenna deployment limits.

	\subsubsection{Multi-waveguide Novel PASS}
	To achieve flexible and efficient deployment, emerging studies have proposed novel multi-waveguide architectures, such as waveguide division multiple access (WDMA) framework and the segmented waveguide-enabled pinching-antenna system (SWAN). Specifically, the authors of \cite{zhao2025waveguide} first introduced the concept of WDMA, where each user is served by a dedicated allocated waveguide. In this scheme, pinching beamforming is exploited to facilitate near-orthogonality during free-space transmission, thereby effectively mitigating inter-user interference. To optimize performance, a joint power allocation and pinching beamforming framework was formulated in \cite{zhao2025waveguide} to maximize the system sum-rate. Besides, the authors of \cite{11348983} proposed SWAN  architecture, employing multiple short dielectric waveguide segments. To optimize connectivity, three practical operating protocols, namely segment selection, segment aggregation, and segment multiplexing, were designed to maximize the uplink and downlink signal-to-noise ratio. Leveraging this segmented architecture, recent research \cite{jiang2025segmented,gan2025revealing} has extended SWAN to explore performance trade-offs in multi-functional integrated systems. Specifically, for integrated sensing and communications (ISAC), the research in \cite{jiang2025segmented} characterized the Pareto fronts of sensing and communication performance and revealed the scaling laws with respect to the number of segments. Furthermore, addressing joint communication and computation, the authors of \cite{gan2025revealing} developed an uplink PASS framework enabled by the segmented-waveguide method. This approach achieves a computation-communication trade-off by supporting the simultaneous transmission of computation data and communication bit streams.

	\subsubsection{Single-waveguide C-PASS}
	Different from multi-waveguide strategies that rely on independent in-waveguide channels, the C-PASS represents a paradigm shift toward single-waveguide architectures, overcoming the $\text{DoF}=1$ bottleneck. Specifically, the C-PASS architecture feeds the input signal into the waveguide via a controllable power splitter, enabling bidirectional signal propagation. This configuration establishes two distinct spatial channels within a single physical waveguide. Consequently, spatial multiplexing gains are unlocked without requiring additional waveguide deployment. To quantify the C-PASS benefits, the authors of~\cite{gan2025c} established the fundamental theoretical framework. By deriving closed-form expressions for DoF and power scaling laws, this work analytically proved that C-PASS doubles the available DoF compared to conventional designs, achieving a multiplexing gain of order $\mathcal{O}(P_{T}\ln^{4}N/N^{2})$, where $P_T$ and $N$ denote the transmit power and the number of pinching elements, respectively. To further facilitate practical deployment of C-PASS, the study in \cite{gan2026center} investigated three operating protocols: power splitting, direction switching, and time switching. To maximize the achievable sum-rate under these protocols, the authors developed tailored optimization algorithms, utilizing weighted minimum mean square error (WMMSE) and penalty-based methods to address the highly coupled constraints involving power allocation and pinching beamforming.

	\subsection{Motivations and Contributions}
	Most of the existing research contributions for DoF enhancement in PASS-assisted communications have concentrated on multi-waveguide architectures. However, linearly increasing the number of waveguides to acquire DoF gains inevitably incurs prohibitive hardware overhead and deployment complexity. Therefore, achieving spatial multiplexing within a single-waveguide architecture represents a significant yet challenging objective. Driven by this motivation, the novel C-PASS architecture was proposed to unlock the spatial potential of a single waveguide. However, current investigations into C-PASS are preliminary, with the achievable DoF strictly limited to $\text{DoF}=2$. Despite its promise, such restricted spatial multiplexing capability is inadequate to meet the stringent spectral efficiency and massive connectivity requirements for B6G networks. This limitation arises because existing C-PASS designs rely on co-located input-port feeding, which provides only two guided propagation directions to distinguish signals.

	To fully exploit the potential of C-PASS, in this paper, we propose a generalized C-PASS framework designed to break the limited DoF constraint. By distributing input ports along the waveguide, the proposed framework creates distinct bidirectional guided paths for different input ports, thereby removing the two-direction bottleneck. More particularly, we derive closed-form expressions for the achievable DoF and power scaling laws. Based on the proposed C-PASS modeling, a joint transmit and pinching beamforming optimization algorithm is developed to maximize the system sum-rate. The main contributions are summarized as follows:	
	\begin{itemize}
		\item We propose a more general framework for C-PASS architecture. Within this framework, we derive the performance analysis for the DoF and power scaling law of the proposed C-PASS. These theoretical results reveal that the C-PASS can achieve $\text{DoF} = \min\{M,K\}$ and the power gain of $\mathcal{O}(P_T M)$, where $M$ and $K$ denote the number of input ports and user antennas, respectively, and $P_T$ denotes the transmit power.
		
		\item We consider a C-PASS empowered MIMO downlink communications, where one single-waveguide C-PASS serves multiple users simultaneously. In particular, we formulate a joint optimization problem for designing the transmit and pinching beamforming to maximize the multi-user sum rate.
		
		\item To solve this highly coupled and non-convex problem, we first transform it into a tractable form via the WMMSE reformulation and develop an efficient alternating optimization algorithm. In this alternating optimization framework, we update the transmit precoding and power splitting ratios through closed-form solutions, and update PA positions and radiation coefficients via block coordinate descent methods.
		
		\item We provide comprehensive numerical results to validate the performance advantages of C-PASS and the effectiveness of the proposed algorithm. The results demonstrate that: 1) the C-PASS significantly enhances the DoF and power scaling law compared to the conventional PASS; 2) For multi-user communications, single-waveguide C-PASS can even outperform multi-waveguide PASS, especially with the high-attenuation waveguide and yield improvement over $10$ dB.

	\end{itemize}

	\subsection{Organization and Notation}
	The remainder of this paper is structured as follows. Section~\ref{sec:model} provides the modeling of the proposed C-PASS architecture and the considered system. Section~\ref{sec:analysis} derives the theoretical performance analysis for DoF and power scaling law of the proposed C-PASS. Section~\ref{sec:design} first formulates the sum-rate maximization problem for C-PASS-assisted downlink multi-user communications. To solve this, section~\ref{sec:design} proposes the joint transmit and pinching beamforming algorithms in an alternating optimization framework. Numerical results evaluate the proposed C-PASS performance and compare it with baselines in Section~\ref{sec:simulation}. Finally, Section \ref{sec:conclusion} concludes the paper.

	\emph{Notations:} Scalars, vectors, and matrices are denoted by italic letters, boldface lowercase letters, and boldface uppercase letters, respectively. $\mathbb{R}^{M \times N}$ and $\mathbb{C}^{M \times N}$ represent the sets of $M \times N$ real and complex matrices. For a matrix $\mathbf{A}$, the superscripts $(\cdot)^T$, $(\cdot)^H$, and $(\cdot)^{-1}$ denote its transpose, Hermitian transpose, and inverse, respectively, while $\text{Tr}(\mathbf{A})$ and $[\mathbf{A}]_{m,n}$ indicate its trace and the $(m,n)$-th entry. $\mathbf{I}_N$ denotes the $N \times N$ identity matrix, and $\text{blkdiag}\{\mathbf{x}_1, \dots, \mathbf{x}_M\}$ constructs a block diagonal matrix with diagonal elements $\mathbf{x}_1, \dots, \mathbf{x}_M$. Moreover, $\|\cdot\|$ and $|\cdot|$ denote the Euclidean norm of a vector and the absolute value of a scalar, respectively. The operator $\odot$ stands for the Hadamard product. Finally, $\mathbb{E}[\cdot]$ and $\Re\{\cdot\}$ denote the statistical expectation and the real part of a complex argument, respectively.

	\section{System Model}\label{sec:model}
	In this section, we present the basic signal model for the proposed C-PASS architecture. Then, we provide the system model for the considered C-PASS-aided communication systems.

	\begin{table}[t]
		\centering
		\caption{Key notation used in the C-PASS signal model.}
		\label{tab:notation}
		
		\begin{tabular}{p{0.25\linewidth}p{0.63\linewidth}}
			\toprule
			Symbol & Definition \\
			\midrule
			$\mathcal{M}_M$ & Index set of the $M$ input ports \\
			$\chi\in\{\mathrm{F},\mathrm{B}\}$ & Propagation direction, forward or backward \\
			$\mathcal{N}_{m,r}^{\chi}$ & Set of PA elements between input port $m$ and region $r$ in direction $\chi$ \\
			$\beta_m^{\mathrm{F}},\beta_m^{\mathrm{B}}$ & Forward and backward power-splitting ratios at input port $m$ \\
			$\delta_n$ & Radiation coefficient of the $n$-th PA \\
			$d_{m,n}^{\mathrm{IN}}$ & In-waveguide distance from input port $m$ to PA region $n$ \\
			$d_{k,n}^{\mathrm{FR}}$ & Free-space distance from PA region $n$ to user $k$ \\
			$\mathbf{Q}$, $\mathbf{H}$ & In-waveguide response matrix and free-space channel matrix \\
			\bottomrule
		\end{tabular}
	\end{table}
	
	\subsection{Proposed C-PASS Architecture}\label{signal_CPASS}
	
	\begin{figure*}[t]
		\centering
		\includegraphics[width=0.7\linewidth]{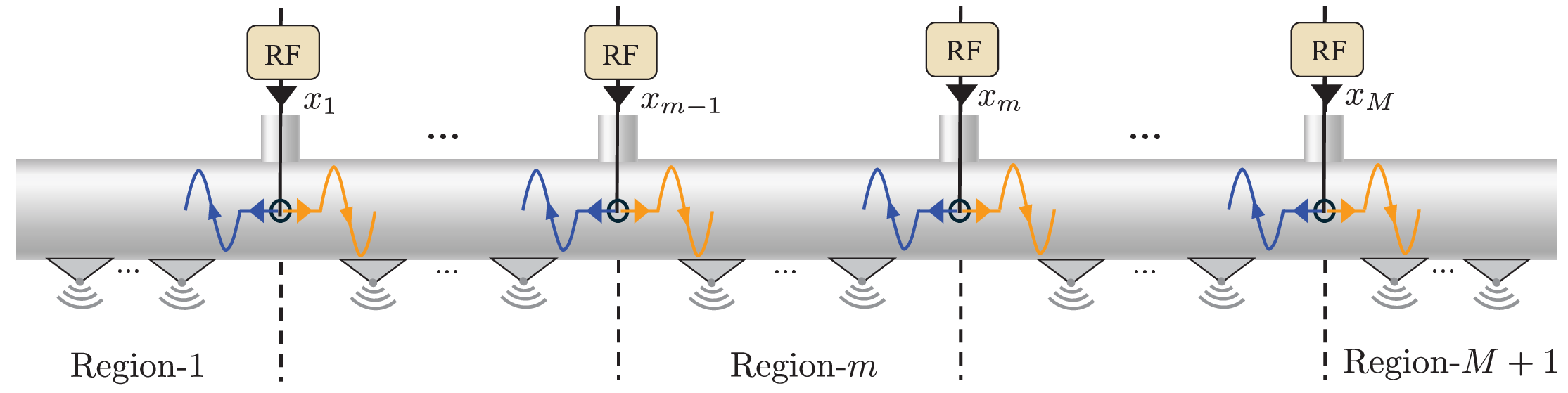}
		\caption{Illustration of a C-PASS architecture.}
		\label{fig:C-PASS}
	\end{figure*}
	
	As shown in Fig.~\ref{fig:C-PASS}, the precoded signal via the radio frequency (RF) is center-fed into the waveguide and divided into a forward-propagation (FP) and backward-propagation (BP) signal, respectively denoted by the yellow and blue lines. To characterize this C-PASS architecture, let $x_m$ denote the signal fed into the $m$-th input port, $m \in \mathcal{M}_M = \{1,2,\dots,M\}$. Based on this, the FP and BP signals at the $m$-th input port can be modeled as\cite{gan2025c}:
	\begin{subequations}
		\begin{align}
			& x_m^{\text{F}} = \sqrt{\beta_m^{\text{F}}} x_m, \\
			& x_m^{\text{B}} = \sqrt{\beta_m^{\text{B}}} x_m,
		\end{align}
	\end{subequations} 
	where the power-splitting ratios satisfy
	\begin{equation}
		\beta_m^{\text{F}}+\beta_m^{\text{B}}=1,
	\end{equation}
	by the law of energy conservation. This power-splitting model follows a calibrated matched traveling-wave assumption, where each input port is sufficiently impedance-matched and the dominant in-waveguide fields are represented by forward and backward traveling waves. Under this communication-level model, $\beta_m^{\mathrm{F}}$ and $\beta_m^{\mathrm{B}}$ denote the effective delivered power ratios after feeding-network calibration, while residual coupling, reflections, and standing-wave effects among multiple input ports can be incorporated as perturbations of the effective in-waveguide channel.
	
	Then, we divide the waveguide into $M+1$ regions based on the position of the input port as the reference point, as shown in Fig.~\ref{fig:C-PASS}. In the $m$-th region ($m \in \mathcal{M}_{M+1}$), there are $N_m$ PAs deployed to radiate communication signals. The in-waveguide propagation can be modeled by the channel~\cite{cheng1989field}: $g =\exp\left( -(\alpha_g+j k_g) d \right)$, where $\alpha_g$ and $k_g$ represent the attenuation and wavenumber of the in-waveguide propagation, respectively. Here, we consider the effective channel from the $m$-th input port to the $n_{m'}$-th FP PA in the $m'$-th region and to the $n_{m''}$-th BP PA in the $m''$-region, respectively for $m'>m$ and $m'' \le m$ as
	\begin{subequations}
		\begin{align}
			& q_{m,n_{m'}}^{\text{F}} = \sqrt{\beta_m^{\text{F}} \xi^{\text{F}}_{m,n_{m'}} } \exp\left( -(\alpha_g+j k_g) d_{m,n_{m'}}^{\text{IN}} \right), \\
			& q_{m,n_{m''}}^{\text{B}} = \sqrt{\beta_m^{\text{B}} \xi^{\text{B}}_{m,n_{m''}}} \exp\left( -(\alpha_g+j k_g) d_{m,n_{m''}}^{\text{IN}}  \right), 
		\end{align}
	\end{subequations}
	where $d^{\text{IN}}_{m,n}$ represents the distance from the $m$-th input port to the $n$-th PA. The power radiation coefficients $\xi^{\text{F}}_{m,n_{m'}}$ and $\xi^{\text{B}}_{m,n_{m''}}$ are utilized to represent the total power radiation ratio from the $m$-th input port to the $n_{m'}$-th FP PA and to the $n_{m''}$-th BP PA, respectively as
	\begin{subequations}
		\begin{align}
			& \xi^{\text{F}}_{m,n_{m'}} =  \prod_{n_1 \in \mathcal{N}_{m,m'}^{\text{F}} } (1-\delta_{n_1}) \prod_{n_2=1}^{n_{m'}-1} (1-\delta_{n_2}) \cdot \delta_{n_{m'}} ,\\
			&\xi^{\text{B}}_{m,n_{m''}} = \prod_{n_1 \in \mathcal{N}_{m,m''}^{\text{B}} } (1-\delta_{n_1}) \prod_{n_2=n_{m''}+1}^{N_{m''}} (1-\delta_{n_2}) \cdot \delta_{n_{m''}},
		\end{align}
	\end{subequations}
	Based on coupled-mode theory of PA~\cite{wang2025modeling}, when the signal $x$ travels through the $n$-th PA, it radiates $\sqrt{\delta_n} x$ into free space, while $\sqrt{1-\delta_n} x$ remains propagating within the waveguide. As can be observed, the expressions for $ \xi^{\chi}_{m,n_{r}}$ can be divided into three parts, respectively accounting for the remaining power from the $m$-th input port to the $r$-th region, remaining power in the $r$-th region, and the radiating power at the $n_r$-th PA.

	\subsection{Signal Model}
	\begin{figure}[t]
		\centering
		\includegraphics[width=0.95\linewidth]{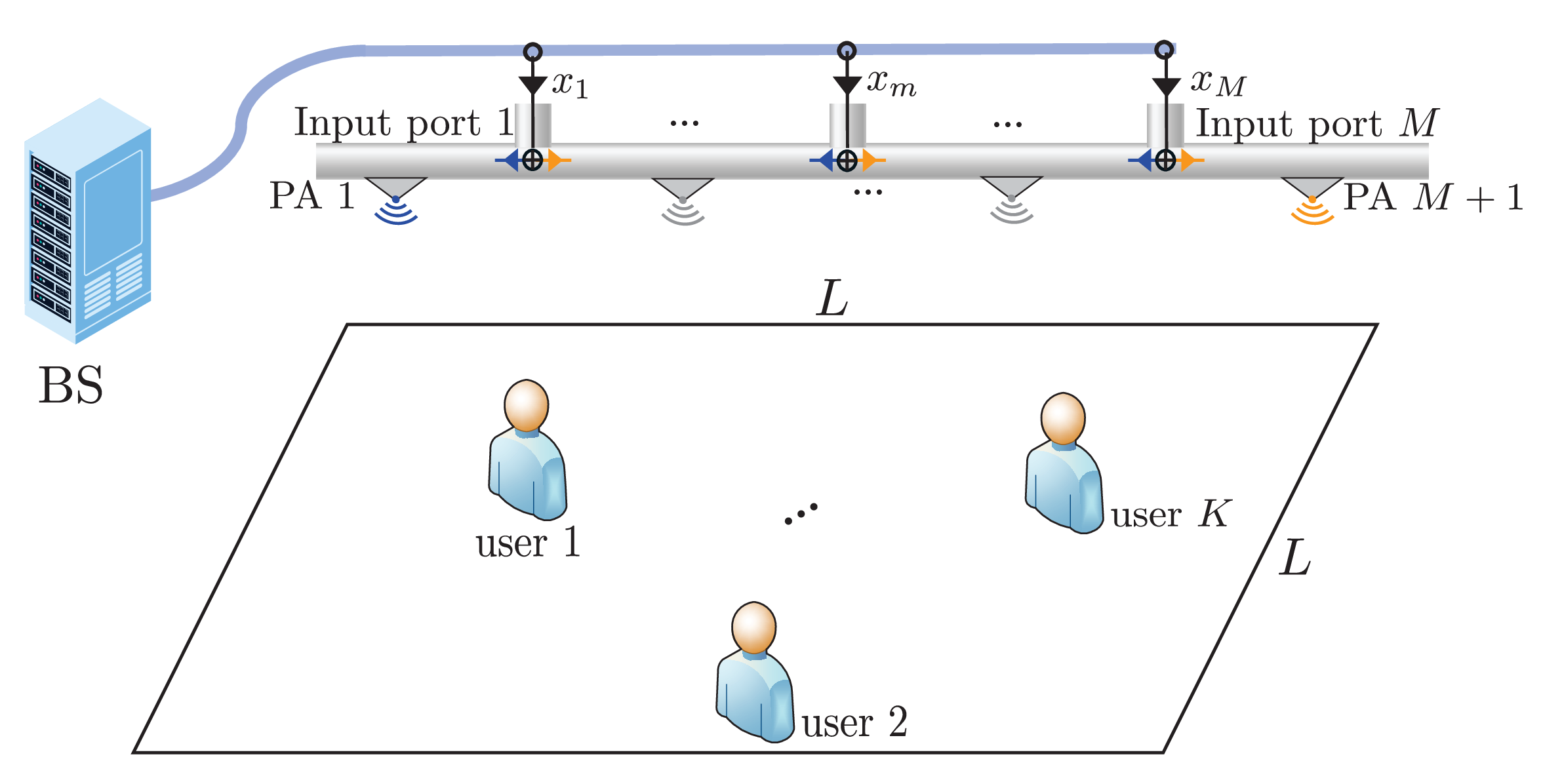}
		\caption{Illustration of a C-PASS-assisted multiuser downlink.}
		\label{fig:system}
	\end{figure}
	
	As illustrated in Fig.~\ref{fig:system}, we consider a base station (BS) serving $K$ users by center-feeding signals in the waveguide through $M$ input ports. Let $\mathbf{P}_m^{\text{IN}}$, $\mathbf{P}_n^{\text{PA}}$ and $\mathbf{P}_k^{\text{US}}$ denote the position coordinates of the $m$-th input port, the $n$-th PA, and the $k$-th user. More particularly, the fed signal $\mathbf{x}=[x_1,x_2,\dots,x_M]^T$ is processed based on the transmit precoding, i.e., $\mathbf{x}=\mathbf{W} \mathbf{s}$, where $\mathbf{s} \in \mathbb{C}^{K \times 1}$ is the information-bearing symbol for $K$ users, and $\mathbf{W} \in \mathbb{C}^{M \times K}$ is the transmit precoding matrix satisfying $\|\mathbf{W}\|^2 \le P_T$ with $P_T$ being the maximum transmit power. Then, the $m$-th fed signal $x_m$ is split into two signals, i.e., $\sqrt{\beta_m^{\text{F}}} x_m$ and $\sqrt{\beta_m^{\text{B}}} x_m$, respectively for the FP and BP directions. It can be assumed that there is only one PA in each region, i.e., $N_m=1$, $\forall m$. Based on this, let $\mathbf{G} \in \mathbb{C}^{M \times (M+1)}$ and $\mathbf{H} \in \mathbb{C}^{K \times (M+1) }$ denote the channel from input ports to the PAs and from the PAs to the users, where the elements of $\mathbf{G}$ and $\mathbf{H}$ can be expressed as
	\begin{subequations}\label{eq:channel1}
		\begin{align}
			& g_{m,n} = \exp\left( -(\alpha_g+j k_g) d_{m,n}^{\text{IN}}  \right), \\
			& h_{k,n} = \eta \exp \left( -j k_0 d_{k,n}^{\text{FR}}  \right)/ d_{k,n}^{\text{FR}},
		\end{align}
	\end{subequations}
	where $d_{m,n}^{\text{IN}} = \left| \mathbf{P}_m^{\text{IN}} - \mathbf{P}_n^{\text{PA}}  \right|$ and $d_{k,n}^{\text{FR}} = \left| \mathbf{P}_k^{\text{US}} - \mathbf{P}_n^{\text{PA}}  \right|$. Specifically, the input ports and PA are assumed to be uniformly deployed along the waveguide. Thus, the distance between the $m$-th input port and $n$-th PA is given by $d_{m,n}^{\text{IN}} = \frac{L}{M+1}\left| m-n + \frac{1}{2} \right|$. $k_0$ is the free-space wavenumber, satisfying $k_0 = 2\pi / \lambda_0$ and $k_0 = k_g/n_{\text{eff}}$, where $n_{\text{eff}}$ is the refractive index of the waveguide. Then, the received signal at the $k$-th user is expressed as
	\begin{equation}
		y_k = \sum_{m=1}^M h_{k,m}^{\text{eff}} x_m + n_k,
	\end{equation} 
	where $ h_{k,m}^{\text{eff}}$ is the effective channel from the $m$-th input port to the $k$-th user, and $n_k \sim \mathcal{CN}(0,N_0)$ denotes the additive noise. Based on the signal model in Section \ref{signal_CPASS}, the effective channel is given by
	\begin{equation}
		\begin{aligned}
			h_{k,m}^{\text{eff}}  = &\sum_{n=1}^m  \sqrt{\beta_m^{\text{B}} \xi^{\text{B}}_{m,n} } g_{m,n} h_{k,n} \\
			& + \sum_{n=m+1}^{M+1} \sqrt{\beta_m^{\text{F}} \xi^{\text{F}}_{m,n} } g_{m,n} h_{k,n},
		\end{aligned}
	\end{equation} 
	which comprises two portions of radiation signals in BP and FP directions. Here, the power radiation coefficients are expressed as
	\begin{subequations}\label{eq:sigma_delta}
		\begin{align}
			& \xi_{m,n}^{\text{B}} = \delta_n \prod_{i=n+1}^m (1-\delta_i), \\
			& \xi_{m,n}^{\text{F}} = \delta_n \prod_{j=m+1}^{n-1} (1-\delta_j).
		\end{align}
	\end{subequations}
	To obtain more compact matrix expression of the effective channel $\mathbf{H}_{\text{eff}} \in \mathbb{C}^{K \times M}$, we have
	\begin{equation}
		\mathbf{H}_{\text{eff}} = \mathbf{H} \mathbf{Q}^T,
	\end{equation}
	where
	\begin{equation}\label{eq:channel_Q}
		[\mathbf{Q}]_{m,n} = 
		\begin{cases} 
			\sqrt{\beta_{m}^{\text{F}} \xi_{m,n}^{\text{F}}} g_{m,n}, & \text{if } n > m, \\
			\sqrt{\beta_{m}^{\text{B}} \xi_{m,n}^{\text{B}}}g_{m,n}, & \text{if } n \le m,
		\end{cases}
	\end{equation}
	for $m \in \{ 1,\dots,M\}$ and $n \in \{1,\dots, M+1 \}$.
	
	\section{Performance Analysis of C-PASS}\label{sec:analysis}
	In this section, the communication performance of the proposed C-PASS is investigated based on the modeling in Section.~\ref{sec:model}. In particular, we derive the DoF and power scaling law in C-PASS, which theoretically verifies the capability of C-PASS to support efficient communications.
	
	\subsection{Degrees of Freedom}\label{sec:DoF}
	To derive the DoF, we first analyze the fundamental capacity limit of the proposed C-PASS architecture. Specifically, here, the system capacity is defined as the maximum mutual information under equal power allocation, expressed as
	\begin{equation}
		\mathcal{C}=\log_2\det\!\left(\mathbf{I}_K+\frac{P_T}{M N_0}\mathbf{H}_{\mathrm{eff}}\mathbf{H}_{\mathrm{eff}}^H\right).
	\end{equation}
	Based on this, the achievable DoF is expressed as
	\begin{equation}\label{eq:DoF}
		\text{DoF} = \lim\limits_{P_T \to \infty} \frac{\mathcal{C}}{\log_2(P_T/N_0)}.
	\end{equation}
	We consider a symmetric configuration where $\beta_m^{\text{F}} = \beta_m^{\text{B}}=1/2$ and $\delta_m=1/2$ for $\forall m$. Based on this configuration, the following theorem formally establishes the DoF achievable by the proposed C-PASS architecture.
	\begin{theorem}
		For the considered symmetric configuration, the achievable DoF of the proposed C-PASS is $\text{DoF}=\min\{M,K\}$.
		\begin{proof}
			Please refer to Appendix~\ref{proof_theorem1}.
		\end{proof}
	\end{theorem}
	\begin{proposition}\label{prop:general_rank}
		Consider the generalized C-PASS with arbitrary power-splitting ratios and radiation coefficients. If $\beta_m^{\mathrm{B}}>0$ and $\delta_m>0$, $\forall m$, and $1-\sqrt{\beta_m^{\mathrm{F}}/\beta_m^{\mathrm{B}}}\sqrt{1-\delta_{m+1}}\rho\neq 0$, $\forall m<M$, where $\rho=\exp\!\left(-(\alpha_g+jk_g)L/(M+1)\right)$, then $\mathbf{Q}$ is full row rank, or equivalently $\mathbf{Q}^T$ is full column rank, and the achievable DoF is $\min\{M,K\}$.
		\begin{proof}
			Please refer to Appendix~\ref{proof_prop_general_rank}.
		\end{proof}
	\end{proposition}
	Proposition~\ref{prop:general_rank} indicates that non-uniform power-splitting ratios, non-uniform radiation coefficients, and constrained PA micro-adjustment may perturb finite-SNR performance, but do not degrade the DoF as long as the effective in-waveguide channel remains non-degenerate. Furthermore, practical mismatch, coupling, and residual reflections mainly perturb the effective in-waveguide channel and may reduce the finite-SNR rate, but they do not remove the rank-based DoF gain unless they make $\mathbf{Q}$ rank deficient. Eq.~\eqref{eq:DoF} explicitly indicates that the DoF determines the asymptotic slope of the channel capacity curve versus $P_T$ in the high power regime. Specifically, in a user-dense scenario where $K > M$, the capacity is fundamentally limited by the number of input ports. Hence, increasing $M$ provides a linear scaling of the sum rate by introducing additional DoF. On the other hand, when $M \ge K$, the C-PASS is capable of the simultaneous service of all $K$ users while effectively utilizing the spatial multiplexing to suppress inter-user interference.
	
	\begin{remark}
		Fundamentally, the C-PASS architecture introduces a new degree of design freedom for the waveguide feeding structure. As such, the achievable spatial DoF is determined by the geometrical distribution of signal input ports and PAs. Specifically, prior studies \cite{gan2025c,gan2026center} primarily focus on a ``co-located input-port feeding'' topology. In this setup, input ports are spatially co-located at the waveguide center, while PAs are deployed on both sides of the feeding point. Although this compact configuration simplifies deployment, the resulting high channel correlation strictly limits the system to a $\text{DoF}=2$. Conversely, the generalized framework proposed in this paper employs a ``distributed input-port feeding'' topology. Here, input ports are spatially spaced along the waveguide, with PAs allocated in the intervals between adjacent input ports. By breaking the correlation structure, the proposed generalized C-PASS design enables high-order spatial multiplexing where the DoF scales linearly with the number of input ports, $M$. From a physical perspective, distributed input-port feeding makes different input ports observe different forward/backward guided distances and phase rotations toward the PA regions. Therefore, the corresponding rows of $\mathbf{Q}$ become distinguishable rather than repeated copies of the same two bidirectional responses in the co-located input-port C-PASS. This explains the motivation and practical advantage of the generalized framework: high-order multiplexing can be realized within a single waveguide, reducing the need for multiple parallel waveguides while supporting multi-user streams and interference suppression.
	\end{remark}
	
	\subsection{Power Scaling Law}
	We then investigate the power scaling law for the received signal in the single-user scenario, i.e., $K=1$. In this case, the optimal transmit precoding adopts the maximum ratio transmission (MRT), i.e., $\mathbf{w} = \sqrt{P_T} \frac{ \mathbf{Q}^* \mathbf{h}^H }{\| \mathbf{Q}^* \mathbf{h}^H \|}$. Thus, the received power under the considered symmetric configuration in Section~\ref{sec:DoF} can be expressed as
	\begin{equation}\label{eq:PSL}
		\begin{aligned}
			& P_R  = P_T \| \mathbf{h}^T \mathbf{Q}^T \|^2 \\
			&=  \frac{P_T}{2} \exp\big( \frac{L\alpha_g }{ M\!+\!1} \big)  \! \sum_{m=1}^{M} \Bigg| \sum_{n=1}^m h_n \varpi^{m-n+1} \!\! +\!\! \!\sum_{n=m+1}^{M+1}  h_n \varpi^{n-m}  \Bigg|^2 \\
			& =   \frac{P_T \eta^2}{2} \exp\big( \frac{L\alpha_g }{ M+1} \big)  \sum_{m=1}^{M}  \Bigg|  \sum_{n=1}^m \frac{\exp(-\frac{L\alpha_g }{ M\!+\!1} (m-n+1) )}{ 2^{(m-n+1)/2} d_{n}^{\text{FR}} } \\
			& \times \exp\left( -j k_0 d_n^{\text{FR}} -j k_g \frac{L (m\!-\!n\!+\!1)  }{ M\!+\!1}\right)  \\
			& + \!\! \sum_{n=m+1}^{M+1}\!\!  \frac{\exp(-\frac{L\alpha_g }{ M\!+\!1} (n\!-\!m) )}{ 2^{(n-m)/2} d_{n}^{\text{FR}} } \exp\!\!\left( \!\!-j k_0 d_n^{\text{FR}}\! -\! j k_g \frac{L (n\!-\!m) }{ M\!+\!1} \right) \! \Bigg|^2.
		\end{aligned} 
	\end{equation}
	By micro-adjusting the positions of PAs to align the phase of each term, the received power is upper-bounded by
	\begin{equation}\label{eq:bound_PSL}
		\begin{aligned}
			\bar{P}_R 
			& = \frac{P_T \eta^2}{2} \exp\big( \frac{L\alpha_g }{ M+1} \big)  \sum_{m=1}^{M}  \Bigg|  \sum_{n=1}^m \frac{\exp(-\frac{L\alpha_g }{ M\!+\!1} (m-n+1) )}{ 2^{(m-n+1)/2} d_{n}^{\text{FR}} } \\
			& + \sum_{n=m+1}^{M+1}\!\!  \frac{\exp(-\frac{L\alpha_g }{ M\!+\!1} (n-m) )}{ 2^{(n-m)/2} d_{n}^{\text{FR}} }  \Bigg|^2.
		\end{aligned} 
	\end{equation}
	\begin{theorem}
		For the considered symmetric configuration and the micro-adjustment of PAs, the power scaling law of the proposed C-PASS is on the order of $\mathcal{O}(P_T M)$.
		\begin{proof}
			Please refer to Appendix~\ref{proof_theorem2}.
		\end{proof}
	\end{theorem}
	
	The insights behind \emph{Theorem 2} reveal that with $N_m=1$ for $\forall m$, the C-PASS behaves similarly to a distributed antenna system with $M$ active antennas, providing a linear power gain of order $\mathcal{O}(M)$. This scaling essentially captures the gain from joint transmission across multiple input ports in C-PASS. More specifically, the distributed input ports create $M$ spatially distinguishable bidirectional in-waveguide responses, which is the architectural source of the high DoF in Theorem~1. This high-DoF structure provides the multiplexing gain that allows the transmitter to jointly exploit $M$ input-port responses, thereby leading to the received-power scaling law $\mathcal{O}(P_T M)$. It can be expected that increasing the number of PAs, i.e., $N_m > 1$, will introduce additional array gains from PAs, thereby further exploiting the performance potential of the proposed C-PASS.

	\section{Beamforming Design of C-PASS}\label{sec:design}
	Based on the proposed C-PASS modeling in Section~\ref{sec:model}, in this section, we formulate a sum-rate maximization problem for communications and propose the optimization algorithm of joint transmit and pinching beamforming designs.
	
	\subsection{Problem Formulation}
	As shown in Fig.~\ref{fig:system}, we consider a sum-rate maximization problem for C-PASS aided downlink communications. Specifically, the received signal can be expressed as
	\begin{equation}
		\mathbf{y} = \mathbf{H} \mathbf{Q}^T \mathbf{W} \mathbf{s} + \mathbf{n},
	\end{equation}
	where $\mathbf{H} = [\mathbf{h}_1, \mathbf{h}_2, \dots, \mathbf{h}_K  ]^T$, and the channel expressions are given in \eqref{eq:channel1} and \eqref{eq:channel_Q}. Then, the sum rate of the $k$-th user in C-PASS is given by
	\begin{equation}
		\mathcal{R} = \sum_{k=1}^K  \log_2 \left( 1+ \frac{|\mathbf{h}_k^T \mathbf{Q}^T \mathbf{w}_k|^2}{\sum_{i \neq k} |\mathbf{h}_k^T \mathbf{Q}^T \mathbf{w}_i|^2 + N_0 }\right).
	\end{equation}
	More particularly, the channel $\mathbf{Q}$ can also be rewritten as
	\begin{equation}\label{eq:expression_Q}
		\begin{aligned}
			\mathbf{Q} = &  \text{diag}( \hat{\bm{\beta}}_{\text{F}} ) \bm{\Psi}(-\mathbf{x}_{\text{IN}}) \bm{\Sigma}_{\text{F}} \bm{\Psi}(\mathbf{x}_{\text{PA}}) \\
			&  + \text{diag}( \hat{\bm{\beta}}_{\text{B}} ) \bm{\Psi}(\mathbf{x}_{\text{IN}}) \bm{\Sigma}_{\text{B}} \bm{\Psi}(-\mathbf{x}_{\text{PA}}),
		\end{aligned}
	\end{equation}
	where 
	\begin{align*}
		& \hat{\bm{\beta}}_{\text{F}} = \left[ \sqrt{\beta_1^{\text{F}}}, \sqrt{\beta_2^{\text{F}}},\dots,\sqrt{\beta_M^{\text{F}}}    \right]^T, \\
		& \hat{\bm{\beta}}_{\text{B}} = \left[ \sqrt{\beta_1^{\text{B}}}, \sqrt{\beta_2^{\text{B}}},\dots,\sqrt{\beta_M^{\text{B}}}    \right]^T, \\
		&  \bm{\Psi}(\mathbf{x}_{\text{PA}}) \!=\! \text{diag}\! \left\{\! \exp(-\!(\!\alpha_g\!+\! j k_g)x_1^{\text{PA}} \!)\!, \dots, \exp(\!-\!(\!\alpha_g\! +\! j k_g\!)\!x_{M\! +\! 1}^{\text{PA}} \!\right\}\!\!, \\
		& \bm{\Psi}(\mathbf{x}_{\text{IN}})\! = \!\text{diag} \!\left\{\! \exp(\!-\!(\alpha_g\!+\!j k_g\!)\!x_1^{\text{IN}} ), \dots, \exp\!(\!-(\alpha_g\!+\!j k_g)x_M^{\text{IN}}\! \right\}\!\!, \\
		& [\bm{\Sigma}_{\text{F}}]_{m,n} = \left\{\begin{matrix}
			\sqrt{\xi_{m,n}^{\text{F}}} 	&  n > m, \\
			0  & \text{otherwise},
		\end{matrix}\right.\\
		& [\bm{\Sigma}_{\text{B}}]_{m,n} = \left\{\begin{matrix}
			\sqrt{\xi_{m,n}^{\text{B}}} 	&  n \le m, \\
			0  & \text{otherwise}.
		\end{matrix}\right.
	\end{align*}
	This indicates that the communication performance is influenced by the design of transmit beamforming: transmit precoding $\mathbf{W}$ and power splitting ratios $\{\hat{\bm{\beta}}_{\text{F}}, \hat{\bm{\beta}}_{\text{B}} \}$, and pinching beamforming: PA position $\mathbf{P}_{\text{PA}}$ and PA power radiation coefficients $\{ \bm{\Sigma}_{\text{F}}, \bm{\Sigma}_{\text{B}} \}$. Thus, a sum-rate maximization problem is formulated to jointly design the transmit and pinching beamforming, given by
	\begin{subequations}\label{pro:ori}
		\begin{align}
			\max_{\mathbf{W},\hat{\bm{\beta}}_{\chi},\mathbf{P}_{\text{PA}},\bm{\Sigma}_{\chi}} \ &   \sum_{k=1}^K \log_2\Big(1+ \frac{|\mathbf{h}_k^T \mathbf{Q}^T \mathbf{w}_k|^2}{\sum_{i \neq k} |\mathbf{h}_k^T \mathbf{Q}^T \mathbf{w}_i|^2 \!+\! N_0 } \Big) \label{pro:ori_obj} \\
			\text{s.t.} \	& \|\mathbf{W}\|^2 \le P_T, \\
			& \beta_m^{\text{F}} + \beta_m^{\text{B}} = 1, \ \forall m, \\
			& \beta_m^{\text{F}}, \beta_m^{\text{B}} \in [0,1] \\
			& \mathbf{P}_{\text{PA}} \in \mathcal{G}(\mathbf{P}_{\text{PA}}), \\
			& \delta_n \in [0,1], \  \forall n,
		\end{align}
	\end{subequations}
	where $\chi \in \{ \text{F}, \text{B}\}$, and $\mathcal{G}(\mathbf{P}_{\text{PA}})$ is the feasible set of all PA positions. Since it is challenging to handle the non-convex log-fractional expression in \eqref{pro:ori_obj}, we use the following lemma to equivalently transform the sum-rate maximization problem into the weighted minimum mean-square error (WMMSE) problem. Specifically, the objective in \eqref{pro:ori_obj} contains logarithmic SINR fractions, where the desired-signal and multi-user interference terms are coupled through $\mathbf{W}$, $\hat{\bm{\beta}}_{\chi}$, PA positions, and radiation coefficients. The WMMSE equivalence introduces receiver and weight auxiliary variables to remove the explicit logarithmic fractional structure. Although the transformed problem remains coupled and non-convex, it yields a weighted MSE form that is more amenable to the subsequent alternating-optimization updates.
	\begin{lemma}
		The objective function~\eqref{pro:ori_obj} can be equivalently transformed into
		\begin{equation}\label{eq:lemma_obj}
			\min_{\bm{\kappa} , \mathbf{t}, \mathbf{W},\hat{\bm{\beta}}_{\chi},\mathbf{P}_{\text{PA}},\bm{\Sigma}_{\chi}} \ \sum_{k=1}^K \kappa_k \epsilon_k - \ln(\kappa_k) ,
		\end{equation} 
		where $\bm{\kappa}=[\kappa_1,\kappa_2,\dots,\kappa_K]^T$, $\bm{\epsilon}=[\epsilon_1,\epsilon_2,\dots,\epsilon_K]^T$ and $\mathbf{t}=[t_1,t_2,\dots,t_K]^T$ are the auxiliary vector, and the equivalent error is defined as
		\begin{equation}
			\epsilon_k = |t_k|^2 \Big(\sum_{i = 1}^K |\mathbf{h}_k^T \mathbf{Q}^T \mathbf{w}_i|^2 + N_0 \Big) - 2\Re\{ t_k^* \mathbf{h}_k^T \mathbf{Q}^T \mathbf{w}_k \} +1.
		\end{equation}
		\begin{proof}
			Please refer to the proof derivation of [\emph{Theorem 1},~\citen{shi2011iteratively}].
		\end{proof}
	\end{lemma} 
	Since the objective function~\eqref{eq:lemma_obj} is convex w.r.t. $t_k$ and $\kappa_k$, the auxiliary variables can be optimized through first-order optimality conditions. Specifically, taking the derivative of $\epsilon_k$ with respect to $t_k^*$ and minimizing $\kappa_k\epsilon_k-\ln(\kappa_k)$ with respect to $\kappa_k$ yield the optimal solution as
	\begin{subequations}\label{update_t_kappa}
		\begin{align}
			& t_k^{\text{opt}} = \frac{\mathbf{h}_k^T \mathbf{Q}^T \mathbf{w}_k }{\sum_{i = 1}^K |\mathbf{h}_k^T \mathbf{Q}^T \mathbf{w}_i|^2 + N_0 }, \\
			& \kappa_k^{\text{opt}} = \epsilon_k^{-1}.
		\end{align}
	\end{subequations} 

	By substituting the derived optimal $t_k^{\text{opt}}$ and $\kappa_k^{\text{opt}}$ into \eqref{eq:lemma_obj}, the optimization problem can be written as
	\begin{subequations}\label{pro:wmmse}
		\begin{align}
			\min_{\mathbf{W},\hat{\bm{\beta}}_{\chi},\mathbf{P}_{\text{PA}},\bm{\Sigma}_{\chi}} \ &  \mathcal{L} (\mathbf{W},\hat{\bm{\beta}}_{\chi},\mathbf{P}_{\text{PA}},\bm{\Sigma}_{\chi}) \label{eq:obj_wmmse}\\
			\text{s.t.} \	& \|\mathbf{W}\|^2 \le P_T, \\
			& \beta_m^{\text{F}} + \beta_m^{\text{B}} = 1, \ \forall m, \\
			& \beta_m^{\text{F}}, \beta_m^{\text{B}} \in [0,1] \\
			& \mathbf{P}_{\text{PA}} \in \mathcal{G}(\mathbf{P}_{\text{PA}}), \\
			& \delta_n \in [0,1], \  \forall n,
		\end{align}
	\end{subequations}
	where the objective function is given by
	\begin{equation}
		\begin{aligned}
			& \mathcal{L} (\mathbf{W},\hat{\bm{\beta}}_{\chi},\mathbf{P}_{\text{PA}},\bm{\Sigma}_{\chi}) \\
			&=  \sum_{k=1}^K 
			\kappa_k \! \Big[|t_k|^2 \sum_{i = 1}^K |\mathbf{h}_k^T \mathbf{Q}^T \mathbf{w}_i|^2 \! -\!  2\Re\{ t_k^* \mathbf{h}_k^T \mathbf{Q}^T \mathbf{w}_k \} \! \Big].
		\end{aligned}
	\end{equation}
	
	\subsection{Subproblem w.r.t. $\mathbf{W}$}\label{algorithm:W}
	With fixed $\{ \hat{\bm{\beta}},\mathbf{P}_{\text{PA}}, \bm{\Sigma}_{\chi} \}$, $\mathbf{H}$ and $\mathbf{G}$ are constant matrix not to optimize. Then, the subproblem w.r.t. $\mathbf{W}$ can be expressed as
	\begin{subequations}
		\begin{align}
			\min_{\mathbf{W}} \ &   \sum_{k=1}^K 
			\kappa_k \! \left[ \! |t_k|^2 \sum_{i = 1}^K \mathbf{w}_i^H \mathbf{Q}^H \mathbf{h}_k^*  \mathbf{h}_k^T \mathbf{Q} \mathbf{w}_i  \! -\!  2\Re\{ t_k^* \mathbf{h}_k^T \mathbf{Q} \mathbf{w}_k \} \! \right]  \\
			\text{s.t.} \	& \|\mathbf{W}\|^2 \le P_T. \label{eq:power_con}
		\end{align}
	\end{subequations}
	By using the Lagrangian function on the power constraint~\eqref{eq:power_con}, the problem is transformed into
	\begin{equation}
		\begin{aligned}
			\min_{\mathbf{W}} \ &   \sum_{k=1}^K 
			\kappa_k \! \left[ \! |t_k|^2 \sum_{i = 1}^K \mathbf{w}_i^H \mathbf{Q}^H \mathbf{h}_k^*  \mathbf{h}_k^T \mathbf{Q} \mathbf{w}_i  \! -\!  2\Re\{ t_k^* \mathbf{h}_k^T \mathbf{Q} \mathbf{w}_k \} \! \right]  \\
			&  + \lambda \left( \sum_{k=1}^K \|\mathbf{w}_k\|^2 - P_T \right).
		\end{aligned}
	\end{equation}
	It can be observed that this unconstrained problem can be solved by setting its first-order derivative to zero, yielding
	\begin{equation}\label{update_W}
		\mathbf{w}_k = \left[ \sum_{i=1}^K \kappa_i |t_i|^2 \mathbf{Q}^H \mathbf{h}_i^*  \mathbf{h}_i^T \mathbf{Q} + \lambda \mathbf{I}_M   \right]^{-1} \kappa_k t_k \mathbf{Q}^H \mathbf{h}_k^*,
	\end{equation}
	where the optimal $\lambda$ is obtained by bisection to satisfy the condition $\lambda^{\text{opt}} \left( \sum_{k=1}^K \|\mathbf{w}_k\|^2 - P_T \right)  = 0$.
	
	\subsection{Subproblem w.r.t. \texorpdfstring{$\hat{\bm{\beta}}_{\chi}$}{beta}}
	With fixed $\{ \mathbf{W}, \mathbf{P}_{\text{PA}}, \bm{\Sigma}_{\chi} \}$, the subproblem w.r.t. the power splitting ratios $\hat{\bm{\beta}}_{\chi}$ can be formulated as
	\begin{subequations}\label{pro:beta}
		\begin{align}
			\min_{\hat{\bm{\beta}}_{\chi}} \ & f_{\bm{\beta}}(\hat{\bm{\beta}}_{\chi})  \\
			\text{s.t.} \	& \big([\hat{\bm{\beta}}_{\text{F}}]_m\big)^2 + \big([\hat{\bm{\beta}}_{\text{B}}]_m\big)^2 = 1, \ \forall m, \label{con:beta1} \\
			& [\hat{\bm{\beta}}_{\text{F}}]_m, [\hat{\bm{\beta}}_{\text{B}}]_m \in [0,1], \ \forall m. \label{con:beta2} 
		\end{align}
	\end{subequations}
	By substituting the expression~\eqref{eq:expression_Q} of $\mathbf{Q}$ into the objective function of \eqref{pro:wmmse}, $f_{\bm{\beta}}(\hat{\bm{\beta}}_{\chi})$ can be expressed as
	\begin{equation}\label{eq:f_beta_obj}
		\begin{aligned}
			& f_{\bm{\beta}}(\hat{\bm{\beta}}_{\chi}) \!=\!\!  \sum_{k=1}^K 
			\kappa_k \! \Big[ \! |t_k|^2 \sum_{i = 1}^K \Big|\mathbf{h}_k^T \big(\bm{\Psi}(\mathbf{x}_{\text{PA}}) \bm{\Sigma}_{\text{F}}^T \bm{\Psi}(-\mathbf{x}_{\text{IN}})  \text{diag}(\mathbf{w}_i) \hat{\bm{\beta}}_{\text{F}} \\
			& + \bm{\Psi}(-\mathbf{x}_{\text{PA}}) \bm{\Sigma}_{\text{B}}^T \bm{\Psi}(\mathbf{x}_{\text{IN}})  \text{diag}(\mathbf{w}_i)  \hat{\bm{\beta}}_{\text{B}}\big)  \Big|^2 \! -\!  2\Re\Big\{ t_k^* \mathbf{h}_k^T \big( \bm{\Psi}(\mathbf{x}_{\text{PA}}) \\
			& \bm{\Sigma}_{\text{F}}^T \bm{\Psi}(-\mathbf{x}_{\text{IN}})  \text{diag}(\mathbf{w}_k)\hat{\bm{\beta}}_{\text{F}} \!+\! \bm{\Psi}(-\mathbf{x}_{\text{PA}}) \bm{\Sigma}_{\text{B}}^T \bm{\Psi}(\mathbf{x}_{\text{IN}})  \text{diag}(\mathbf{w}_k) \hat{\bm{\beta}}_{\text{B}} \big) \!\! \Big\} \! \Big].
		\end{aligned}
	\end{equation}
	
	To tackle the non-convexity arising from the coupled constraint~\eqref{con:beta1}, we adopt the block coordinate descent (BCD) method to iteratively optimize the variables $\{[\hat{\bm{\beta}}_{\text{F}}]_m, [\hat{\bm{\beta}}_{\text{B}}]_m \}$ for the $m$-th block while keeping others fixed. Specifically, we utilize a trigonometric parameterization to automatically satisfy the constraints \eqref{con:beta1} and \eqref{con:beta2}:
	\begin{equation}
		[\hat{\bm{\beta}}_{\text{F}}]_m = \cos(\theta_m), \quad [\hat{\bm{\beta}}_{\text{B}}]_m = \sin(\theta_m),
	\end{equation}
	where $\theta_m \in [0,\frac{\pi}{2}]$, $\forall m$. Consequently, the objective function~\eqref{eq:f_beta_obj} w.r.t. $\theta_m$ can be rewritten as
	\begin{equation}\label{eq:f_theta_obj}
		\begin{aligned}
			& f_{\bm{\theta}}(\theta_m) \!=\!\!  \sum_{k=1}^K \!
			\kappa_k \! \Big[  |t_k|^2\!  \sum_{i = 1}^K\! \Big| [\bm{\eta}_{k,i}^{\text{F}}]_m \cos(\theta_m)\! +\! [\bm{\eta}_{k,i}^{\text{B}}]_m \sin(\theta_m)\!  \\
			& \!+\! C_{k,i}^{-m}  \Big|^2 \!\!\! - \!2\Re\big\{ \! t_k^* ([\bm{\eta}_{k,k}^{\text{F}}]_m\! \cos(\theta_m)\! +\! [\bm{\eta}_{k,k}^{\text{B}}]_m \!\sin(\theta_m)\!  +\! C_{k,k}^{-m} ) \!  \big\}\! \Big], 
		\end{aligned}
	\end{equation}
	where the coefficients are defined as
	\begin{subequations}
		\begin{align}
			& \bm{\eta}_{k,i}^{\text{F}} = \mathbf{h}_k^T \bm{\Psi}(\mathbf{x}_{\text{PA}}) \bm{\Sigma}_{\text{F}}^T \bm{\Psi}(-\mathbf{x}_{\text{IN}})  \text{diag}(\mathbf{w}_i), \\
			& \bm{\eta}_{k,i}^{\text{B}} = \mathbf{h}_k^T  \bm{\Psi}(-\mathbf{x}_{\text{PA}}) \bm{\Sigma}_{\text{B}}^T \bm{\Psi}(\mathbf{x}_{\text{IN}})  \text{diag}(\mathbf{w}_i) .
		\end{align}
	\end{subequations}
	The constant term collecting the coefficient values from other blocks is denoted by
	\begin{equation}
		C_{k,i}^{-m} = \sum_{m' \neq m} ([\bm{\eta}_{k,i}^{\text{F}}]_{m'} \cos(\theta_{m'})\! +\! [\bm{\eta}_{k,i}^{\text{B}}]_{m'} \sin(\theta_{m'})).
	\end{equation} 
	Since the objective function~\eqref{eq:f_theta_obj} is continuously differentiable w.r.t. $\theta_m$ and bounded within $[0,\frac{\pi}{2}]$, the optimal $\theta_m$ resides either at a stationary point satisfying the first-order optimality conditions or at the boundary\cite{nocedal2006numerical}. By setting the first-order derivative of \eqref{eq:f_theta_obj} to zero, we obtain
	\begin{equation}
		A_m \sin(2 \theta_m) + B_m \cos(2 \theta_m) +C_m \sin(\theta_m) +D_m  \cos(\theta_m) = 0,
	\end{equation} 
	where the coefficients are given by
	\begin{align*}
		& A_m =  \sum_{k=1}^K 
		\kappa_k \!  |t_k|^2\!  \sum_{i = 1}^K ( |[\bm{\eta}_{k,i}^{\text{B}}]_m |^2 -|[\bm{\eta}_{k,i}^{\text{F}}]_m |^2 ), \\
		& B_m =  2 \sum_{k=1}^K 
		\kappa_k \!  |t_k|^2\!  \sum_{i = 1}^K \Re\big\{  [\bm{\eta}_{k,i}^{\text{F}}]_m^* [\bm{\eta}_{k,i}^{\text{B}}]_m    \big\}, \\
		& C_m =\!  2 \!\sum_{k=1}^K\! 
		\kappa_k\! \Big[- \! |t_k|^2\!  \sum_{i = 1}^K\! \Re\big\{ \! [\bm{\eta}_{k,i}^{\text{F}}]_m^* C_{k,i}^{-m}  \big\}\! +\!  \Re\big\{ t_k^* [\bm{\eta}_{k,k}^{\text{F}}]_m \big\}\!\Big], \\
		& D_m =\! 2\! \sum_{k=1}^K \!
		\kappa_k \!\Big[  |t_k|^2\!  \sum_{i = 1}^K \Re\big\{ \! [\bm{\eta}_{k,i}^{\text{B}}]_m^* C_{k,i}^{-m} \! \big\}\! -\!  \Re\big\{ t_k^* [\bm{\eta}_{k,k}^{\text{B}}]_m \big\}\! \Big].
	\end{align*}
	To unify the trigonometric terms into a single variable, we employ the Weierstrass substitution with $z_m=\tan(\frac{\theta_m}{2}) \in [0,1]$. Using the identities $\sin(\theta_m)=\frac{2z_m}{1+z_m^2}$, $\cos(\theta_m) = \frac{1-z_m^2}{1+z_m^2}$, $\sin(2\theta_m) = \frac{4z_m(1-z_m^2)}{(1+z_m^2)^2}$, and $\cos(2\theta_m) = \frac{1-6z_m^2+z_m^4}{(1+z_m^2)^2}$, the equation $\frac{\partial f_{\bm{\theta}}}{\partial \theta_m} = 0$ is transformed into
	\begin{equation}\label{pro:zm}
		\begin{aligned}
			\underbrace{(B_m-D_m)}_{c_{m,4}} z_m^4 + \underbrace{(-4 A_m + 2C_m)}_{c_{m,3}} z_m^3 + \underbrace{(- 6 B_m)}_{c_{m,2}} z_m^2 \\
			+ \underbrace{(4 A_m + 2C_m)}_{c_{m,1}} z_m + \underbrace{ B_m + D_m}_{c_{m,0}} = 0.
		\end{aligned}
	\end{equation}
	Utilizing Ferrari's method, the four potential roots of \eqref{pro:zm} are derived as
	\begin{subequations}\label{eqs:zm}
		\begin{align}
			& z_{m,1} = -\frac{c_{m,3}}{4c_{m,4}} + S_m + \frac{1}{2}\sqrt{-4 S_m^2 - 2 p_m  - \frac{q_m}{S_m}}, \\
			& z_{m,2} = -\frac{c_{m,3}}{4c_{m,4}} + S_m - \frac{1}{2}\sqrt{-4 S_m^2 - 2 p_m  - \frac{q_m}{S_m}}, \\
			& z_{m,3} = -\frac{c_{m,3}}{4c_{m,4}} - S_m + \frac{1}{2}\sqrt{-4 S_m^2 - 2 p_m  + \frac{q_m}{S_m}}, \\
			& z_{m,4} = -\frac{c_{m,3}}{4c_{m,4}} - S_m - \frac{1}{2}\sqrt{-4 S_m^2 - 2 p_m  + \frac{q_m}{S_m}},
		\end{align}
	\end{subequations}
	where
	\begin{align*}
		& p_m = (8c_{m,4} c_{m,2} - 3c_{m,3}^2)/(8c_{m,4}^2) \\
		& q_m = (c_{m,3}^3 - 4c_{m,4} c_{m,3}c_{m,2} + 8c_{m,4}^2 c_{m,1})/(8c_{m,4}^3), \\
		& S_m = \frac{1}{2} \sqrt{-\frac{2}{3}p_m + \frac{1}{3c_{m,4}} \left( Q_m + \frac{\Delta_{m,0}}{Q_m} \right)}, \\
		& Q_m = \sqrt[3]{\big(\Delta_{m,1} + \sqrt{\Delta_{m,1}^2 - 4\Delta_{m,0}^3}\big)/2}, \\
		& \Delta_{m,0} = c_{m,2}^2 - 3c_{m,3} c_{m,1} + 12c_{m,4} c_{m,0}, \\
		& \Delta_{m,1} = 2c_{m,2}^3 - 9c_{m,3} c_{m,2} c_{m,1} + 27c_{m,3}^2 c_{m,0} + 27c_{m,4} c_{m,1}^2 \\
		&- 72c_{m,4} c_{m,2} c_{m,0}.
	\end{align*}
	Thus, the set of candidate solutions is given by
	\begin{equation}
		\mathcal{Z}_m \!=\! \left\{ z \!\in\! \left\{ z_{m,1}, z_{m,2}, z_{m,3}, z_{m,4} \right\}\! \mid\! z\in[0,1], z\in \mathbb{R} \right\}\! \cup\! \{0,1\}.
	\end{equation}
	The above filtering follows from $z_m=\tan(\theta_m/2)$ and $\theta_m\in[0,\pi/2]$, which imply $z_m\in[0,1]$. Hence, complex roots and real roots outside $[0,1]$ are discarded before the minimization step.
	Finally, the optimal $\theta_m$ is determined by selecting the candidate that minimizes the objective function:
	\begin{equation}
		\theta_m^{\text{opt}} = \arg \min_{z \in \mathcal{Z}_m} f_{\bm{\theta}}(2 \arctan(z)).
	\end{equation}
	After updating $\theta_m^{\text{opt}}$ for all blocks until convergence, the designed $\hat{\bm{\beta}}_{\chi}$ is obtained by
	\begin{equation}\label{update_hat_beta}
		[\hat{\bm{\beta}}_{\text{F}}]_m = \cos(\theta_m^{\text{opt}}), \quad [\hat{\bm{\beta}}_{\text{B}}]_m = \sin(\theta_m^{\text{opt}}).
	\end{equation}

	\subsection{Subproblem w.r.t. $\mathbf{P}_{\text{PA}}$}
	In the subproblem w.r.t. the PA positions, we consider a linear deployment where the PAs are arranged along the horizontal axis. Let $\mathbf{X}_{\text{PA}} = [X_1^{\text{PA}}, \dots, X_{M+1}^{\text{PA}}]^T$ denote the $X$-axis position coordinate vector of PAs to be optimized. Considering the physical hardware implementation of the PASS, large-scale displacement of PAs is impractical. Consequently, the position of the $m$-th PA is constrained within a micro-adjustment interval $[x_m^{\text{PA,init}} - \Delta, x_m^{\text{PA,init}} + \Delta]$, where $x_m^{\text{PA,init}}$ is the initial position coordinate of the $m$-th PA, and $\Delta$ represents the maximum deviation of the micro-adjustment. Even in this setting, the WMMSE objective in \eqref{pro:wmmse} remains highly coupled and non-convex w.r.t. $\mathbf{X}_{\text{PA}}$. To solve this, we adopt the BCD method that updates one block of PA positions at a time while keeping the others fixed. In this way, the optimization of the PA positions can be decoupled into $M+1$ subproblems.
	
	More particularly, we investigate the impact of the position coordinate $X_m^{\text{PA}}$ on the $m$-th column of $\mathbf{H}$, denoted as $\mathbf{h}_m(X_m^{\text{PA}})$, and the $m$-th row of $\mathbf{Q}$, denoted as $\mathbf{q}_m^T(X_m^{\text{PA}})$. 
	\begin{subequations}\label{eq:pa_bcd_subproblem}
		\begin{align}
			\min_{X_m} \quad & \sum_{k=1}^{K}\kappa_k\Big[
			|t_k|^2\,\big\|\mathbf{e}_k^T\big(\mathbf{C}_{-m}^{\text{PA}}+\mathbf{h}_m(X_m)\mathbf{q}_m^T(X_m)\big)\mathbf{W}\big\|^2 \notag \\
			&-2\Re\!\big\{t_k^*\,\mathbf{e}_k^T\big(\mathbf{C}_{-m}^{\text{PA}}+\mathbf{h}_m(X_m)\mathbf{q}_m^T(X_m)\big)\mathbf{w}_k\big\}
			\Big] \label{eq:pa_bcd_obj}\\
			\text{s.t.} \quad & |X_m-X_m^{\text{PA,init}}|\le \Delta, \label{eq:pa_bcd_tr}
		\end{align}
	\end{subequations}
	where $\mathbf{C}_{-m}^{\text{PA}} = \sum_{i=1,\,i \neq m}^{M+1} \mathbf{h}_i(X_i^{\text{PA}})\mathbf{q}_i^T(X_i^{\text{PA}})$ collects all the contributions of all PAs except $m$ to $\mathbf{H}_{\text{eff}}$ and is constant w.r.t. $X_m$ for this subproblem~\eqref{eq:pa_bcd_subproblem}. To address this, we adopt a high-resolution one-dimensional grid search strategy. By discretizing the feasible interval into $N_{\text{grid}}$ candidate points, we evaluate and select the optimal position that minimizes the WMMSE objective~\eqref{eq:pa_bcd_obj}. This procedure is performed sequentially for all $m=1, \dots, M+1$, ensuring a monotonic decrease of the objective function value until convergence.

	\subsection{Subproblem w.r.t. \texorpdfstring{$\bm{\Sigma}_{\chi}$}{beta}}\label{algorithm:sigma}
	With fixed $\{\mathbf{W}, \hat{\bm{\beta}}_{\chi}, \mathbf{P}_{\text{PA}}\}$, the subproblem w.r.t. $\bm{\Sigma}_{\text{F}}$ and $\bm{\Sigma}_{\text{B}}$ is jointly determined by the PA power splitting ratios $\bm{\delta}=[\delta_1,\dots,\delta_{M+1}]^T$ in \eqref{eq:sigma_delta}. This induces the subproblem w.r.t. $\bm{\Sigma}_{\chi}$ can be transformed into the optimization of $\bm{\delta}$. Note that the power radiation ratio of the endpoint PAs is fixed as $\delta_1 = \delta_{M+1} = 1$ to radiate all the power of the remaining signals. However, the optimization of $\bm{\delta}$ in the proposed C-PASS architecture presents two challenges as follows. (1) Due to the center-fed input ports excitation, $\bm{\Sigma}_{\chi}$ evolve into lower-triangular structures rather than the simple diagonal forms in conventional PASS. (2) Two directional radiation matrices $\bm{\Sigma}_{\text{F}}$ and $\bm{\Sigma}_{\text{B}}$ are high-order functions of the PA radiation ratio vector $\bm{\delta}$ in \eqref{eq:sigma_delta}, and are intrinsically coupled. As a result, the $\bm{\delta}$-subproblem becomes highly coupled and non-convex, which motivates the adoption of BCD-based update. More particularly, we utilize the trigonometric parameterization to represent the expressions for
	\begin{equation}
		\sqrt{\delta_m} = \cos(\phi_m), \quad \sqrt{1-\delta_m} = \sin(\phi_m),
	\end{equation}
	where $\phi_m \in [0,\frac{\pi}{2}]$ for $m \in \{2,\dots,M\}$. Based on this, we employ the overall PA power radiation matrix as
	\begin{equation}
		[\bm{\Sigma}]_{m,n} = \left\{\begin{matrix}
			\cos(\phi_n) \prod_{i=m+1}^{n-1} \sin(\phi_i), 	&  n > m ,\\
			\cos(\phi_n) \prod_{i=n+1}^{m} \sin(\phi_i), 	&  n \le m .
		\end{matrix}\right.
	\end{equation}
	Then, the effective in-waveguide channel can be rewritten as
	\begin{equation}
		\mathbf{Q} = \bm{\Sigma} \odot \mathbf{U},
	\end{equation}
	where the remaining coefficient is given by
	\begin{equation}
		[\mathbf{U}]_{m,n} = \left\{\begin{matrix}
			\sqrt{\beta_{m}^{\text{F}}} g_{m,n}	& n > m , \\
			\sqrt{\beta_{m}^{\text{B}}} g_{m,n} & n \le m .
		\end{matrix}\right.
	\end{equation}
	The subproblem w.r.t. the $n$-th block to update $\{\cos(\phi_n), \sin(\phi_n) \}$ is formulated as
	\begin{subequations}\label{prob:phi_n}
		\begin{align}
			\min_{\phi_n} \quad &  \sum_{k=1}^K \kappa_k \Big[ |t_k|^2 \left\| \mathbf{W}^T \big[ \bm{\Sigma}(\phi_n) \odot \mathbf{U} \big] \mathbf{h}_k \right\|^2 \notag \\
			&\quad - 2\Re\left\{ t_k^* \mathbf{h}_k^T\big[ \bm{\Sigma}(\phi_n) \odot \mathbf{U} \big]^T \mathbf{w}_k \right\} \Big]  \label{eq:phi_n_obj}\\
			\text{s.t.} \quad & 0 \le \phi_n \le \frac{\pi}{2},
		\end{align}
	\end{subequations}
	where $\bm{\Sigma}(\phi_n)$ captures the impact of $\phi_n$ on the PA power radiation matrix while keeping the set $\{\phi_j\}_{j \neq n}$ fixed. It can be observed that the objective function~\eqref{eq:phi_n_obj} is continuously differentiable over the compact feasible set $[0,\pi/2]$. However, deriving a closed-form solution for the stationary point is analytically intractable due to the high-order polynomial nature of the product terms in $\bm{\Sigma}(\phi_n)$. Hence, to efficiently determine the optimal $\phi_n$ in each BCD iteration, we employ Brent's algorithm~ \cite{brent2013algorithms}, which combines golden-section search with inverse parabolic interpolation to minimize the objective function~\eqref{eq:phi_n_obj} over $[0, \pi/2]$ up to a prescribed tolerance, i.e., $10^{-4}$. This sequential update procedure is iterated for $n=2, \dots, M$ until the fractional decrease of the objective function falls below a predefined threshold. Upon convergence, the optimized power splitting ratios are recovered from the phase variables as
	\begin{equation}
		[\bm{\delta}]_n = \cos^2(\phi_n), \quad \forall n \in \{2, \dots, M\},
	\end{equation}
	with the boundary elements fixed as $[\bm{\delta}]_1 = [\bm{\delta}]_{M+1} = 1$. Consequently, the power radiation matrices, $\bm{\Sigma}_{\text{F}}$ and $\bm{\Sigma}_{\text{B}}$, are determined according to \eqref{eq:sigma_delta}.
	
	\subsection{Overall Algorithm}
	\begin{algorithm}[t]
		\caption{Proposed Alternating Optimization Algorithm to Maximize the Sum Rate for Solving Problem~\eqref{pro:ori}}
		\label{algorithm:proposed}
		\begin{algorithmic}[1]
			\STATE Initialize the optimization variables.
			\REPEAT
			\STATE Update $\mathbf{t}$ and $\bm{\kappa}$ via closed-form solutions~\eqref{update_t_kappa}.
			\STATE Update $\mathbf{W}$ via closed-form solutions~\eqref{update_W} with $\lambda^{\text{opt}}$.
			\STATE Update $\hat{\bm{\beta}}_{\chi}$ via closed-form solutions~\eqref{update_hat_beta}.
			\STATE BCD-based update $\mathbf{P}_{\text{PA}}$ by solving problems~\eqref{eq:pa_bcd_subproblem} through 1D grid search.
			\STATE BCD-based update $\bm{\Sigma}_{\chi}$ by solving problems~\eqref{prob:phi_n} through Brent's algorithm.
			\UNTIL{Convergence or reaching the maximum number of iterations $I_{\max}$.}
		\end{algorithmic}
	\end{algorithm}
	
	Based on the algorithm design for solving the optimization variables $\{ \mathbf{W},\hat{\bm{\beta}}_{\chi},\mathbf{P}_{\text{PA}},\bm{\Sigma}_{\chi} \}$, the overall alternating optimization algorithm is summarized in Algorithm~\ref{algorithm:proposed}. Specifically, the alternating update process is terminated when the relative increment of the sum-rate falls below a predefined threshold $\epsilon$ or the maximum number of iterations $I_{\max}$ is reached. More particularly, in the $l$-th iteration, the optimization variables are updated sequentially as follows:
	\begin{itemize}
		\item The precoding matrix $\mathbf{W}$ is updated via the Lagrangian-based method to obtain the global minimum, yielding $\mathcal{L}(\mathbf{W}^{l+1}, \hat{\bm{\beta}}_{\chi}^l, \mathbf{P}_{\text{PA}}^l, \bm{\Sigma}_{\chi}^l) \le \mathcal{L}(\mathbf{W}^{l}, \hat{\bm{\beta}}_{\chi}^l, \mathbf{P}_{\text{PA}}^l, \bm{\Sigma}_{\chi}^l)$.
		
		\item The power splitting ratios $\hat{\bm{\beta}}_{\chi}$ are updated by selecting the optimal roots of the derived quartic equation, ensuring $\mathcal{L}(\mathbf{W}^{l+1}, \hat{\bm{\beta}}_{\chi}^{l+1}, \mathbf{P}_{\text{PA}}^l, \bm{\Sigma}_{\chi}^l) \le \mathcal{L}(\mathbf{W}^{l+1}, \hat{\bm{\beta}}_{\chi}^l, \mathbf{P}_{\text{PA}}^l, \bm{\Sigma}_{\chi}^l)$.
		
		\item The PA positions $\mathbf{P}_{\text{PA}}$ are refined via a high-resolution grid search which selects a candidate position that reduces the cost, resulting in $\mathcal{L}(\mathbf{W}^{l+1}, \hat{\bm{\beta}}_{\chi}^{l+1}, \mathbf{P}_{\text{PA}}^{l+1}, \bm{\Sigma}_{\chi}^l) \le \mathcal{L}(\mathbf{W}^{l+1}, \hat{\bm{\beta}}_{\chi}^{l+1}, \mathbf{P}_{\text{PA}}^l, \bm{\Sigma}_{\chi}^l)$.
		
		\item The radiation matrices $\bm{\Sigma}_{\chi}$ are updated by optimizing the phase variables using Brent's method, satisfying $\mathcal{L}(\mathbf{W}^{l+1}, \hat{\bm{\beta}}_{\chi}^{l+1}, \mathbf{P}_{\text{PA}}^{l+1}, \bm{\Sigma}_{\chi}^{l+1}) \le \mathcal{L}(\mathbf{W}^{l+1}, \hat{\bm{\beta}}_{\chi}^{l+1}, \mathbf{P}_{\text{PA}}^{l+1}, \bm{\Sigma}_{\chi}^l)$.
	\end{itemize}
	Since the objective function $\mathcal{L}(\mathbf{W},\hat{\bm{\beta}}_{\chi},\mathbf{P}_{\text{PA}},\bm{\Sigma}_{\chi}) $ is strictly non-increasing throughout the iteration process and is bounded from below, the sequence of objective values is guaranteed to converge. Finally, based on the equivalence between the WMMSE formulation and the sum-rate maximization problem, the proposed algorithm is guaranteed to converge to a stationary point of \eqref{pro:ori_obj}.
	
	To make the computational cost of each update block explicit, Table~\ref{tab:complexity} summarizes the per-outer-iteration complexity of Algorithm~\ref{algorithm:proposed}. The $\mathbf{W}$-subproblem mainly involves matrix inversion and bisection over the Lagrange multiplier. The $\hat{\bm{\beta}}_{\chi}$-subproblem is solved by BCD over $M$ power-splitting blocks. The PA-position and radiation-coefficient updates are also BCD-type procedures, where the dominant costs come from the one-dimensional grid search and Brent's function evaluations, respectively.
	\begin{table}[t]
		\centering
		\caption{Per-outer-iteration computational complexity of Algorithm~\ref{algorithm:proposed}.}
		\label{tab:complexity}
		
		\begin{tabular}{l l}
			\toprule
			Update block & Complexity \\
			\midrule
			$\mathbf{W}$ & $\mathcal{O}\!\left((M^3+KM^2)\log_2(1/\epsilon_{\lambda})\right)$ \\
			$\hat{\bm{\beta}}_{\chi}$ & $\mathcal{O}(I_{\beta}MK^2)$ \\
			$\mathbf{P}_{\mathrm{PA}}$ & $\mathcal{O}(I_{\mathrm{PA}}N_{\mathrm{grid}}MK(M+K))$ \\
			$\bm{\Sigma}_{\chi}$ & $\mathcal{O}(I_{\Sigma}N_{\mathrm{Brent}}M^2K(M+K))$ \\
			\bottomrule
		\end{tabular}
	\end{table}
	Therefore, with $I_{\mathrm{AO}}$ outer iterations, the overall computational complexity of the proposed algorithm is given by $\mathcal{O}\big( I_{\text{AO}} \big[ (M^3 + K M^2) \log_2(1/\epsilon_{\lambda}) + I_{\beta} M K^2 + I_{\text{PA}} N_{\text{grid}} MK(M + K) + I_{\Sigma} M N_{\text{Brent}} MK(M + K) \big] \big)$, where $I_{\text{AO}}$ denotes the number of outer iterations required for the alternating optimization to converge, $\epsilon_{\lambda}$ denotes the accuracy of the bisection search, $I_{\beta}$, $I_{\mathrm{PA}}$, and $I_{\Sigma}$ denote the BCD iteration numbers of the corresponding subproblems, and $N_{\mathrm{grid}}$ and $N_{\mathrm{Brent}}$ denote the number of grid-search candidates and Brent function evaluations, respectively. This complexity expression also suggests practical low-complexity variants. Since the PA-position update and radiation-coefficient update scale with $N_{\mathrm{grid}}$ and $N_{\mathrm{Brent}}$, respectively, one can reduce the implementation cost by using a coarser PA-position grid, fixed PA positions, or fixed radiation coefficients. These variants trade a small pinching-beamforming loss for a lower repeated BCD cost.

	\section{Numerical Results}\label{sec:simulation}
	This section provides numerical simulations validating the derived DoF and power scaling laws for C-PASS in \emph{Theorems 1 and 2}. Furthermore, we evaluate the effectiveness of the proposed algorithms in enhancing the communication performance.
	
	\subsection{Simulation Setup}
	For our simulations, $K$ users are randomly located on a square area of $100 \times 100$ m${}^2$. The C-PASS is deployed along $Y=0$ with the length of $L=100$ m. The $M$ input ports are uniformly fed into the waveguide within $[0,100]$ m, where the $M+1$ PA are deployed in the exact center between the two input ports with micro adjustments. As such, the $X$-axis coordinates of the $m$-th input port and PA are $ \frac{m L}{M+1}$ and $\frac{(m-\frac{1}{2})  L}{M+1}$, respectively for $m \in [1,M]$ and $m \in [1,M+1]$. The positions of PA can be micro-adjusted in a small-scale range of $[-\Delta,\Delta]$ with $\Delta=0.01$ m. The operating frequency band is considered at the mmWave band of $f_c=77$ GHz~\cite{dapa2023vehicular}. Besides, the effective refractive index of the waveguide is set as $n_{\text{eff}}=1.44$. The in-waveguide amplitude attenuation coefficient is assumed as $\alpha=0.0092$~\cite{xu2025pinching}. The free-space channel coefficient is computed as $\eta=\frac{\lambda_0}{4\pi}$. Unless stated otherwise, the simulation setup is given as follows: $K=4$, $M=4$, $P_T=30$ dB, $N_0=-90$ dBm. The convergence parameter for the proposed algorithm is set as $\epsilon=10^{-3}$ and $I_{\max}=100$. For fair comparison, all considered schemes use the same total transmit power budget $P_T$, user locations, carrier frequency, noise power, waveguide attenuation coefficient, path-loss model, convergence threshold, and maximum iteration number.
	
	\subsection{Baseline Schemes}
	To verify the effectiveness of the proposed C-PASS architecture and algorithms, we evaluate and compare with the following schemes and baselines.
	\begin{itemize} 
		\item \textbf{Scheme 1 (PA Position Micro-Tuning Only):} In this scheme, the power splitting ratios and PA radiation coefficients are fixed to constant values, while the transmit precoding is designed via the MRT strategy. Consequently, only the PA positions are refined by applying the proposed \emph{Algorithm 1}.
		
		\item \textbf{Baseline 2 (Conventional Single-Waveguide End-Fed PASS):} This baseline represents a conventional single-waveguide end-fed PASS, where $M$ input ports are placed near the waveguide terminal to feed signals into one waveguide. The PAs are initialized with the same uniformly distributed positions as in the proposed C-PASS. Specifically, the $X$-axis coordinates of the $m$-th input port and the $m$-th PA are given by $X_{m}^{\mathrm{IN}} = \frac{5}{4}\frac{2\pi}{k_g} m$, $m\in[1,M]$, and $X_{m}^{\mathrm{PA}} = \frac{(m-1/2)L}{M+1}$, $m\in[1,M+1]$, respectively. The joint optimization of $\mathbf{W}$, $\mathbf{P}_{\text{PA}}$, and $\bm{\Sigma}_{\chi}$ is performed using \emph{Algorithm 1}, subject to the end-fed constraint $\beta_m^{\text{F}}=1, \beta_m^{\text{B}}=0, \forall m$.
		
		\item \textbf{Baseline 3 (Conventional $M$-Waveguide PASS):} This baseline employs $M$ parallel waveguides, where the $m$-th waveguide is deployed at the spacing $Y_{\text{WG}}=1$ m at a coordinate of $-(m-1) Y_{\text{WG}}$. Each waveguide operates with a single input port and a single PA. For the $m$-th waveguide, the $X$-axis coordinates of the input port and the PA are initialized as $X_m^{\mathrm{IN}}=\frac{5}{4}\frac{2\pi}{k_g} m$ and $X_m^{\mathrm{PA}}=\frac{(m-1/2)L}{M}$, respectively. Similarly, the optimization of $\mathbf{W}$, $\mathbf{P}_{\text{PA}}$, and $\bm{\Sigma}_{\chi}$ is solved via \emph{Algorithm 1} by enforcing the single-direction feeding constraint $\beta_m^{\text{F}}=1, \beta_m^{\text{B}}=0, \forall m$.
	\end{itemize}

	\subsection{Performance Validation of Theoretical Results}
	\begin{figure}[t]
		\centering
		\includegraphics[width=0.95\linewidth]{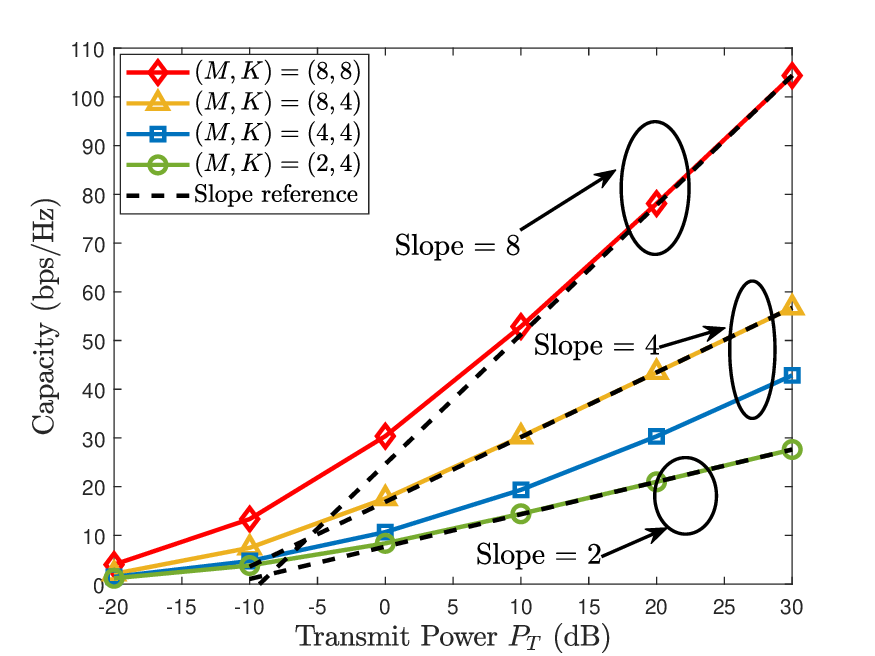}
		\caption{DoF characterization of C-PASS.}
		\label{sim:DoF}
	\end{figure}
	
	\begin{figure}[t]
		\centering
		\includegraphics[width=0.95\linewidth]{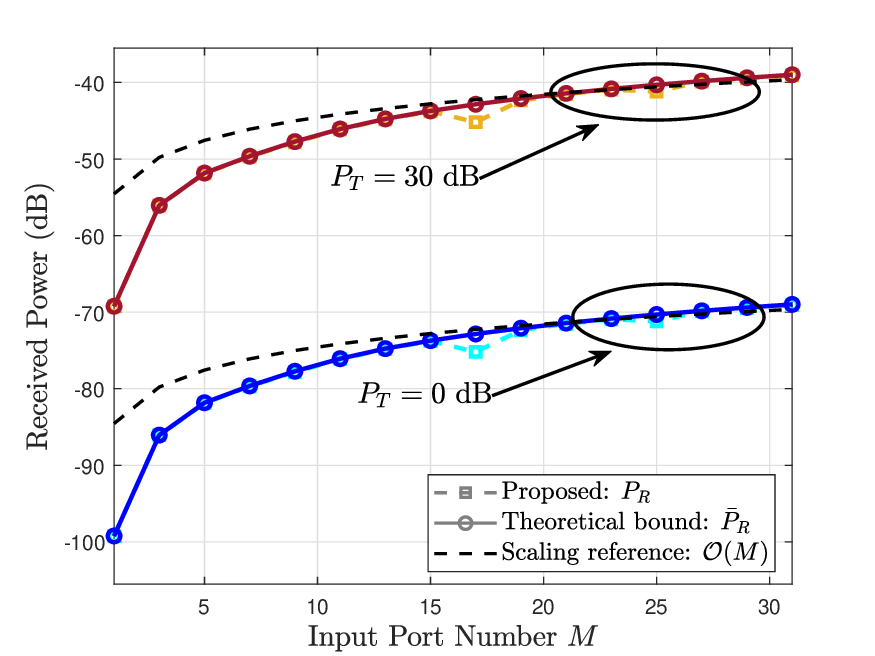}
		\caption{Power scaling law characterization of C-PASS.}
		\label{sim:PSL}
	\end{figure}
	
	In Fig.~\ref{sim:DoF}, we evaluate the DoF of the proposed C-PASS under the configuration of Scheme 1. Specifically, the system parameters are set as $\beta_m^{\text{F}}=\beta_m^{\text{B}}=1/2$ and $\delta_m=1/2$ for $\forall m$. To quantify the DoF under different $M$ and $K$, we examine the slope of the capacity curves in the high transmit power regime, given by Eq.~\eqref{eq:DoF}.	As illustrated in Fig.~\ref{sim:DoF}, in the high-$P_T$ regime, the capacity curve for the configuration $(M,K)=(8,8)$ exhibits a slope parallel to the reference line of $8$. Similarly, the curves corresponding to $(M,K)=(8,4)$ and $(M,K)=(4,4)$ align closely with the reference slope of $4$, while the configuration with $(M,K)=(2,4)$ asymptotically matches the reference slope of $2$. These results indicate that the achievable DoFs for these four configurations are $8$, $4$, $4$, and $2$, respectively, which validates the theoretical derivation of $\text{DoF}=\min\{M,K\}$ in \emph{Theorem 1}. This observation intuitively reveals that by exploiting $M$ input ports to center-feed signals, the C-PASS architecture significantly scales up the DoF performance. Thus, the proposed C-PASS can offer substantial capacity gains to support multi-user MIMO communications, compared with conventional PASS.
	
	In Fig.~\ref{sim:PSL}, we investigate the power scaling law of the received signal power in dB versus the number of input ports, $M$. It follows the configuration of \textbf{Scheme 1}, where $\beta_m^{\text{F}}=\beta_m^{\text{B}}=1/2$ and $\delta_m=1/2$ for all $m$. Specifically, the ``Proposed'' curve represents the actual received power $P_R$ calculated via \eqref{eq:PSL} using the PA position micro-tuning optimization. For comparison, the ``Theoretical bound'' curve plots the analytical upper bound $\bar{P}_R$ derived in \eqref{eq:bound_PSL}. As observed from Fig.~\ref{sim:PSL}, the curves for the proposed scheme and the theoretical bound overlap almost perfectly across the large $M$ for both $P_T=0$ and $30$ dB cases. This tight alignment demonstrates that even within a highly constrained PA movement range (i.e., small $\Delta=0.01$ m), the proposed micro-tuning strategy effectively realizes the theoretical upper bound of the received power. Furthermore, we introduce a scaling reference line representing $\mathcal{O}(M)$, which is plotted as $10\log_{10}(M)$ in the dB domain. It is evident that as $M$ increases, both the proposed and theoretical curves asymptotically converge to this reference line. This observation explicitly validates the theoretical derivation in \emph{Theorem 2}, confirming that the power scaling law of the proposed C-PASS achieves an order of $\mathcal{O}(M)$. It also reveals that increasing the number of input ports not only expands the spatial DoF but also yields a significant multiplexing power gain of C-PASS.

	\subsection{Convergence of the Proposed Algorithm}
	\begin{figure}[t]
		\centering
		\includegraphics[width=0.95\linewidth]{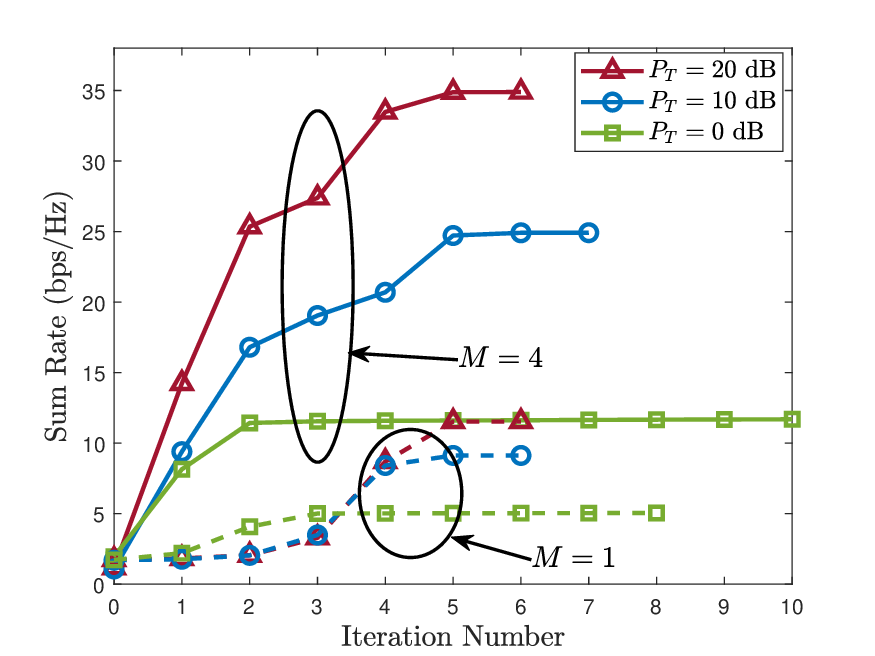}
		\caption{Convergence behavior of the proposed algorithm.}
		\label{sim:convergence}
	\end{figure}
	
	Fig.~\ref{sim:convergence} illustrates the convergence behavior of \emph{Algorithm 1} with different numbers of input ports, $M$, and transmit power, $P_T$. It is observed that for all considered settings, the sum rate monotonically increases with the number of iterations and converges within $10$ iterations. This confirms the fast convergence of the proposed algorithm. To further quantify the runtime behavior, Table~\ref{tab:runtime_convergence} reports the execution time corresponding to the six curves in Fig.~\ref{sim:convergence}. The simulations were conducted in MATLAB R2023b on a macOS 15.5 platform with an Apple M4 processor and 16 GB memory. Furthermore, the performance gain achieved by increasing $M$ from $1$ to $4$ is significantly larger than that obtained by increasing $P_T$ from $0$ dB to $20$ dB. This indicates that increasing the number of input ports is a more energy-efficient approach to improve the system performance compared to increasing the transmit power.
	
	\begin{table}[t]
		\centering
		\caption{Runtime corresponding to the convergence curves in Fig.~\ref{sim:convergence}.}
		\label{tab:runtime_convergence}
		\begin{tabular}{c c c c}
			\toprule
			$M$ & $P_T$ (dB) & Iterations & Runtime (s) \\
			\midrule
			1 & 0  & 8  & 0.0615 \\
			1 & 10 & 6  & 0.0457 \\
			1 & 20 & 6  & 0.0380 \\
			4 & 0  & 10 & 0.2332 \\
			4 & 10 & 7  & 0.1601 \\
			4 & 20 & 6  & 0.1380 \\
			\bottomrule
		\end{tabular}
	\end{table}

	\subsection{Performance Comparison with Baseline Schemes}
	\begin{figure}[t]
		\centering
		\includegraphics[width=0.95\linewidth]{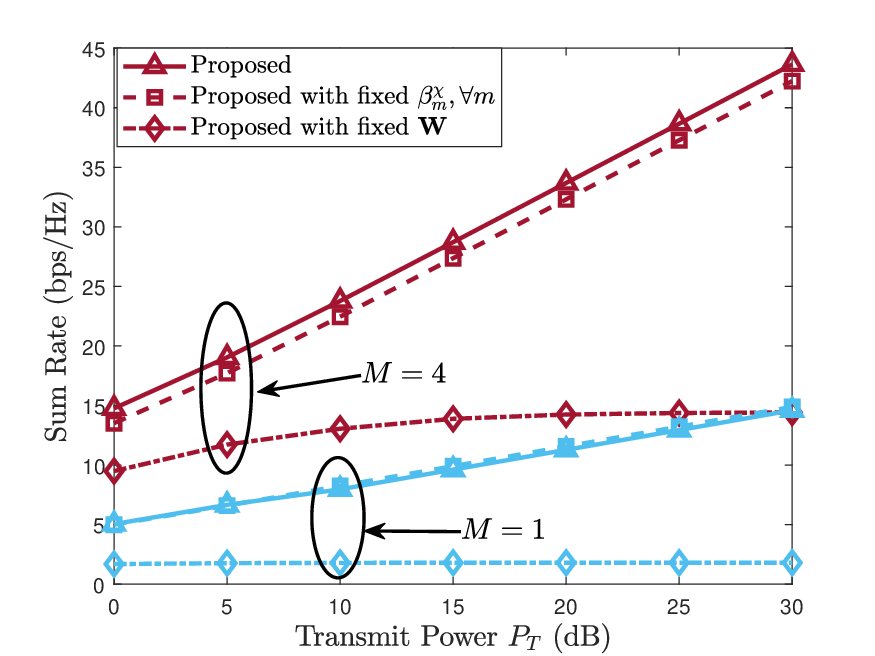}
		\caption{Comparison with transmit-beamforming baselines.}
		\label{sim:benchmark_transmit}
	\end{figure}	
	
	\begin{figure}[t]
		\centering
		\includegraphics[width=0.95\linewidth]{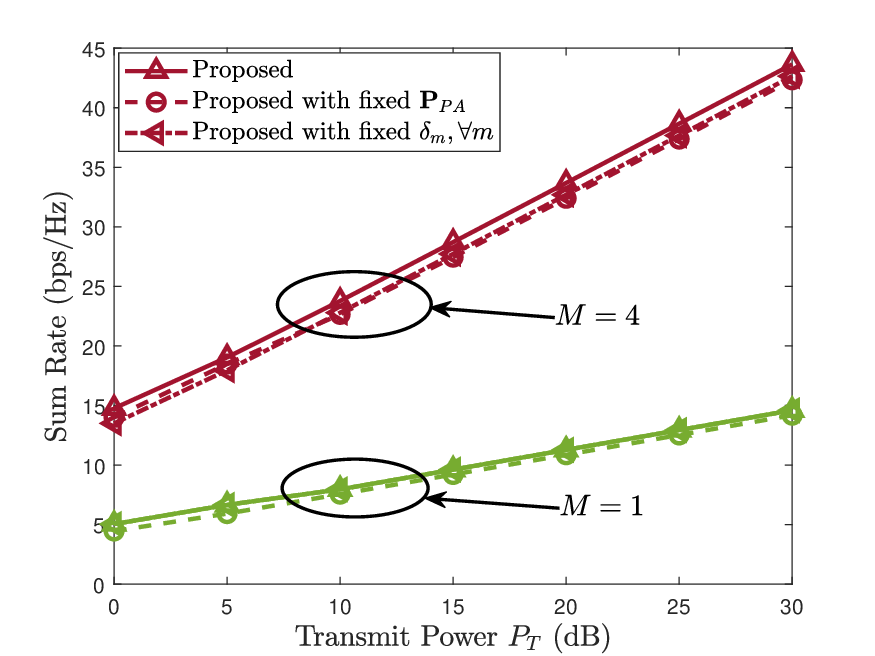}
		\caption{Comparison with pinching-beamforming baselines.}
		\label{sim:benchmark_pinching}
	\end{figure}
	
	In Figs.~\ref{sim:benchmark_transmit} and \ref{sim:benchmark_pinching}, we evaluate the impact of transmit and pinching beamforming designs on the C-PASS performance. Specifically, we compare the proposed \emph{Algorithm 1}, involving the joint optimization of $\{ \mathbf{W},\hat{\bm{\beta}}_{\chi},\mathbf{P}_{\text{PA}},\bm{\Sigma}_{\chi} \}$, with baseline schemes where subsets of these optimization variables are set to fixed values. We first focus on investigating the impact of the transmit beamforming design on C-PASS in Fig.~\ref{sim:benchmark_transmit}. It is observed that employing a fixed power splitting strategy with $\beta_m^{\text{F}}=\beta_m^{\text{B}}=1/2$ for all $m$ incurs only a marginal sum-rate degradation, particularly in the low-$M$ regime. This indicates that non-adjustable power splitters are sufficient to maintain near-optimal performance, thereby offering a practical trade-off between system capacity and hardware complexity. Conversely, fixing the transmit precoding results in a significant performance deterioration, which becomes increasingly pronounced as $M$ and $P_T$ increase. 
	
	In Fig.~\ref{sim:benchmark_pinching}, we further evaluate the impact of the pinching beamforming design. It is observed that fixing the PA positions without applying the position micro-tuning incurs a negligible sum-rate loss. Similarly, setting a constant radiation coefficient of $\delta_m=1/2$ for $m=2,\dots,M$ results in only a marginal performance degradation. This robustness indicates that the transmit beamforming optimization can effectively compensate for the fixed pinching settings. This is because the transmit beamforming adapts to the effective channel established by the fixed PAs, thereby maintaining the system capacity. These results suggest that a simplified pinching beamforming configuration, coupled with optimized transmit beamforming, is sufficient to guarantee satisfactory performance in practical C-PASS. These observations also provide useful insights for practical PA implementations of C-PASS, such as two-state PASS designs where the PA radiation state is selected from a finite set while the PA position is fixed~\cite{11414084}. The limited loss under fixed pinching settings suggests that the main C-PASS gain is expected to be largely preserved with such discrete PA designs.

	\subsection{Performance Comparison with Baseline Architectures}
	\begin{figure}[t]
		\centering
		\includegraphics[width=0.95\linewidth]{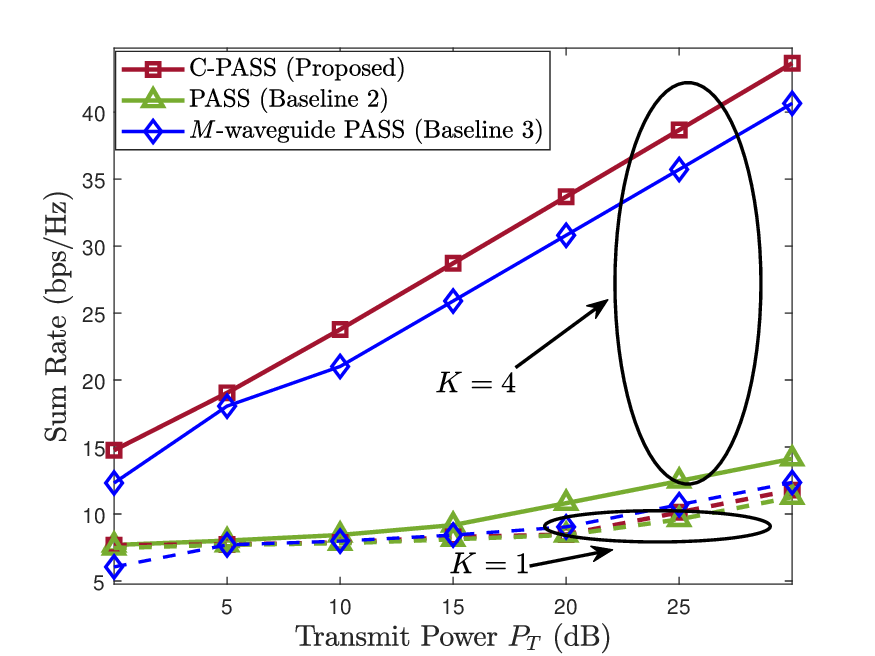}
		\caption{Comparison with conventional PASS architectures.}
		\label{sim:conventional}
	\end{figure}

	\begin{figure}[t]
		\centering
		\includegraphics[width=0.95\linewidth]{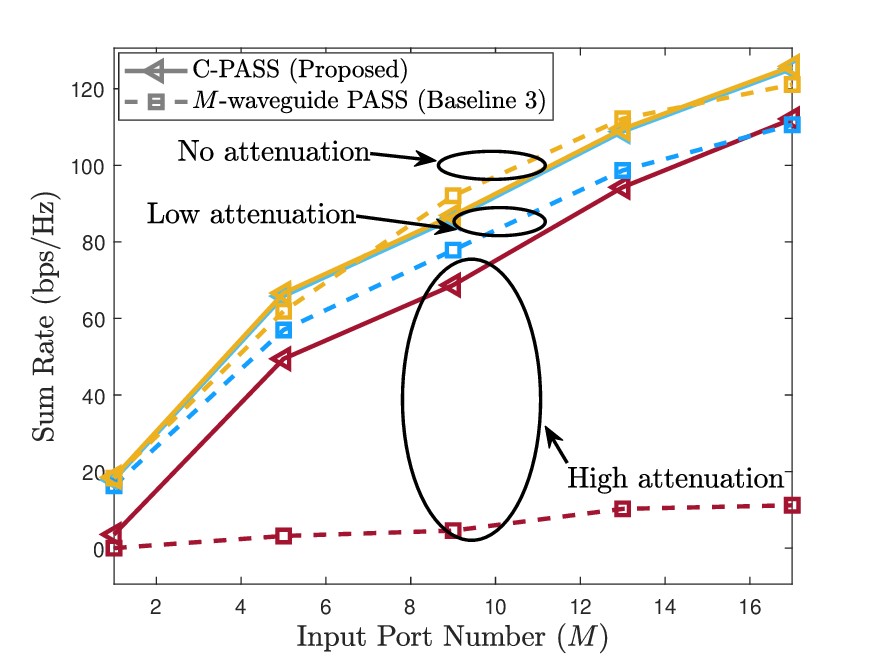}
		\caption{Comparison under different in-waveguide attenuation with $K=10$.}
		\label{sim:attenuation}
	\end{figure}
	
	In Fig.~\ref{sim:conventional}, we evaluate the performance of the proposed C-PASS against Baseline 2 and Baseline 3 under $K=1$ and $K=4$. In the single-user scenario, the performance gap between C-PASS and the two baseline architectures is negligible. This is attributed to the fact that the single spatial DoF provided by the conventional PASS is sufficient to support single-stream transmission, making the additional DoF offered by C-PASS or the multi-waveguide PASS unnecessary for performance enhancement. In contrast, for the multi-user scenario, the conventional single-waveguide PASS exhibits severe performance degradation due to the rank deficiency. To address this limitation, the multi-waveguide PASS effectively expands the spatial multiplexing capability via the physical deployment of multiple waveguide channels. However, signals in the end-fed architecture are highly susceptible to signal power loss due to the long in-waveguide transmission distance, which significantly constrains the sum-rate performance. Conversely, the proposed C-PASS effectively addresses this bottleneck by employing a distributed center-feeding architecture. This design substantially increases DoF and shortens the effective signal propagation path within the waveguide, thereby mitigating the attenuation and achieving superior sum-rate performance.
	
	Fig.~\ref{sim:attenuation} investigates the impact of in-waveguide attenuation by comparing the sum-rate performance of C-PASS and the multi-waveguide PASS under three scenarios: the ideal no-attenuation case with $\alpha=0$~\cite{ding2025flexible}, the low-attenuation case with $\alpha=0.0092$~\cite{xu2025pinching}, and the high-attenuation case with $\alpha=0.2095$~\cite{vu2021experimental}. It is observed that the performance of the multi-waveguide PASS deteriorates significantly as the attenuation coefficient increases, particularly in the large-$M$ regime. This degradation is attributed to the accumulated propagation loss along the long in-waveguide transmission, which becomes a critical bottleneck for the end-fed multi-waveguide architecture. In contrast, C-PASS exhibits remarkable robustness to attenuation. Specifically, the performance degradation in the low-attenuation scenario is negligible compared to the ideal case. Even in the high-attenuation scenario with $M=17$, the C-PASS incurs a loss of only $0.5$ dB relative to the ideal baseline. Notably, in this specific setting, the C-PASS achieves a performance gain of approximately $10$ dB over the multi-waveguide baseline. Therefore, the relative gain of C-PASS becomes more pronounced as the attenuation coefficient increases, since distributed center feeding shortens the effective guided propagation path and avoids the accumulated loss of end-fed architectures.

	\section{Conclusion}\label{sec:conclusion}
	The generalized framework for C-PASS architecture was investigated, aiming to break the fundamental DoF bottleneck. Based on the proposed C-PASS and basic signal modeling, closed-form expressions for the DoF and power scaling laws were derived in \emph{Theorems 1 and 2}, respectively. Then, a joint transmit and pinching beamforming optimization problem was formulated to maximize the system sum rate. More particularly, the resulting highly coupled non-convex problems were efficiently solved by WMMSE reformulation and alternating optimization algorithms. Numerical results validated the analytical derivations, confirming the effectiveness of C-PASS in significantly enhancing communication performance. Furthermore, the obtained results also revealed that the single-waveguide C-PASS is capable of outperforming multi-waveguide architectures, particularly for severe waveguide attenuation.
	
	The results obtained in this paper confirm the effectiveness of employing C-PASS for improving the performance of wireless networks, which motivates related future research. In essence, the center-fed architecture introduces an additional dimension of configuration flexibility to the PASS framework. Specifically, this generalized structure regards the conventional end-fed PASS as a special case, obtained when the input signals are confined to a single propagation direction. Consequently, the performance lower bound of the C-PASS is theoretically guaranteed to match that of the conventional end-fed baselines, ensuring robust performance enhancements for wireless networks. Moreover, the linear DoF scaling with the number of input ports, $M$, established in the single-waveguide C-PASS can be extended to multi-waveguide architectures. Specifically, with the deployment of $G$ waveguides, the system DoF of C-PASS is increased to scale multiplicatively to $M \cdot G$. This multiplicative increase provides abundant spatial resources, thereby facilitating adaptation to complex communication environments and enabling advanced functionalities.

	\appendices
	\section{Proof of Theorem 1}\label{proof_theorem1}
	Following the proof of \emph{Theorem 1} in \cite{gan2025c}, the DoF derivation can be transformed into the rank of the effective channel, given by
	\begin{equation}
		\text{DoF} = \text{rank}\left( \mathbf{H}_{\text{eff}} \right).
	\end{equation}
	The rank of PASS-aided communication is primarily constrained by the in-waveguide channel. Thus, it can be assumed that the channel between random-position users and PAs to be full-rank, i.e., $\text{rank}(\mathbf{H})=\min\{ K, M+1 \}$. In the following, we focus on studying the rank of the effective in-waveguide channel $\mathbf{Q}$ in \eqref{eq:channel_Q}. Based on the considered configuration, the element of $\mathbf{Q}$ can be rewritten as
	\begin{equation}\label{eq:Q_matrix}
		\begin{aligned}
			\mathbf{Q}  = &  \sqrt{\frac{1}{2}} \exp\left( \frac{L(\alpha_g+j k_g) }{ 2 ( M+1)}   \right) \\
			& \times  \begin{bmatrix}
				\varpi & \varpi & \varpi^2 & \cdots &  \varpi^M \\
				\varpi^2 & \varpi & \varpi & \cdots & \varpi^{M-1} \\
				\varpi^3 & \varpi^2 & \varpi & \cdots & \varpi^{M-2} \\
				\vdots & \ddots & \ddots & \ddots & \vdots\\
				\varpi^M & \varpi^{M-1} & \varpi^{M-2} & \cdots & \varpi		
			\end{bmatrix},
		\end{aligned}
	\end{equation}
	where $\varpi = \sqrt{\frac{1}{2}} \exp\left(- (\alpha_g+j k_g) \frac{L}{ M+1}   \right)$. We first investigate the rank of the partial in-waveguide channel matrix, i.e., $\mathbf{Q}_0 = [\mathbf{Q}]_{1:M, 1:M}$. Based on the expression in \eqref{eq:Q_matrix}, the determinant of $\mathbf{Q}_0$ can be expressed as
	\begin{equation}
		\begin{aligned}
			\text{det}(\mathbf{Q}_0) & = \left[ \sqrt{\frac{1}{2}} \exp\left( \frac{L(\alpha_g+j k_g) }{ 2 ( M+1)}   \right) \varpi  \right]^M (1-\varpi)^{M-1} \\
			& \neq 0,
		\end{aligned}
	\end{equation}
	which derives the matrix $\mathbf{Q}^T$ has full column rank. According to the rank factorization property, there exists an invertible transformation matrix $\mathbf{T}_Q \in \mathbb{C}^{(M+1) \times (M+1)}$ such that $\mathbf{Q}^T$ can be transformed into a canonical form:\begin{equation}\label{eq:canonical_form}\mathbf{T}_Q \mathbf{Q}^T = \left[\begin{matrix}
			\mathbf{I}_M \\
			\mathbf{0}_{1 \times M}
		\end{matrix}\right].
	\end{equation}
	Consequently, the effective channel can be rewritten as 
	\begin{equation}
		\begin{aligned}
			\mathbf{H}_{\text{eff}}  & = \mathbf{H}  \left( \mathbf{T}_Q^{-1} \mathbf{T}_Q \right) \mathbf{Q}^T \\
			& = \tilde{\mathbf{H}} \left[\begin{matrix}
				\mathbf{I}_M \\
				\mathbf{0}_{1 \times M}
			\end{matrix}\right] = \left[ \tilde{\mathbf{H}} \right]_{:,1:M},
		\end{aligned}
	\end{equation}
	where $\tilde{\mathbf{H}} = \mathbf{H} \mathbf{T}_Q^{-1}$. Since $\mathbf{T}_Q$ is an invertible matrix for $\mathbf{Q}$, the linear transformation $\tilde{\mathbf{H}} = \mathbf{H} \mathbf{T}_Q^{-1}$ preserves the rank of the original channel matrix $\mathbf{H}$, i.e., 
	\begin{equation}
		\text{rank}(\tilde{\mathbf{H}}) = \min\{ M+1, K\}.
	\end{equation}
	Since $\tilde{\mathbf{H}}$ is a full-rank matrix, its column vectors are linearly independent. Thus, the sub-matrix formed by extracting the first $M$ columns achieves the rank of $\min\{ M ,K\}$. Then, we can obtain
	\begin{equation}
		\text{rank}(\mathbf{H}_{\text{eff}}) = \min \{ K, M \}.
	\end{equation}
	
	\section{Proof of Proposition~\ref{prop:general_rank}}\label{proof_prop_general_rank}
	We extend the rank argument in Appendix~\ref{proof_theorem1} to general coefficients. Let $\rho=\exp\!\left(-(\alpha_g+jk_g)L/(M+1)\right)$, and define $\mathbf{Q}_0=[\mathbf{Q}]_{1:M,1:M}$. Since the $n$-th PA is at $x_n^{\mathrm{PA}}=(n-\frac{1}{2})L/(M+1)$ and the $m$-th input port at $x_m^{\mathrm{IN}}=mL/(M+1)$, the in-waveguide distances satisfy $d_{m,n}^{\mathrm{IN}}=(m-n+\frac{1}{2})L/(M+1)$ for $n\le m$ and $d_{m,n}^{\mathrm{IN}}=(n-m-\frac{1}{2})L/(M+1)$ for $n>m$, giving $g_{m,n}=\rho^{m-n+1/2}$ and $g_{m,n}=\rho^{n-m-1/2}$, respectively. Substituting into \eqref{eq:channel_Q} and \eqref{eq:sigma_delta} yields
	\begin{equation}
		[\mathbf{Q}_0]_{m,n}=
		\begin{cases}
			\rho^{m-n+\frac{1}{2}}\sqrt{\beta_m^{\mathrm{B}}\delta_n}
			\displaystyle\prod_{i=n+1}^{m}\!\sqrt{1-\delta_i}, & n\le m,\\[6pt]
			\rho^{n-m-\frac{1}{2}}\sqrt{\beta_m^{\mathrm{F}}\delta_n}
			\displaystyle\prod_{i=m+1}^{n-1}\!\sqrt{1-\delta_i}, & n>m.
		\end{cases}
	\end{equation}
	Since $\beta_m^{\mathrm{B}}>0$ and $\delta_n>0$, we factor $\sqrt{\beta_m^{\mathrm{B}}}$ from row $m$ and $\rho^{1/2}\sqrt{\delta_n}$ from column $n$, writing
	\begin{equation}
		\det(\mathbf{Q}_0)=\rho^{M/2}\prod_{m=1}^{M}\sqrt{\beta_m^{\mathrm{B}}}\prod_{n=1}^{M}\sqrt{\delta_n}\cdot\det(\mathbf{A}),
	\end{equation}
	where the cleaned matrix $\mathbf{A}\in\mathbb{C}^{M\times M}$ has entries (with $r_k=\sqrt{1-\delta_k}\,\rho$ and $a_m=\sqrt{\beta_m^{\mathrm{F}}/\beta_m^{\mathrm{B}}}$; empty products equal one)
	\begin{equation}\label{eq:A_entries}
		[\mathbf{A}]_{m,n}=
		\begin{cases}
			\displaystyle\prod_{i=n+1}^{m}r_i, & n\le m,\\[6pt]
			\displaystyle a_m\prod_{i=m+1}^{n-1}r_i, & n>m.
		\end{cases}
	\end{equation}
	We evaluate $\det(\mathbf{A})$ by considering its leading $k\times k$ submatrix $\mathbf{A}_k$ and applying the elementary row operation $R_k\leftarrow R_k-r_kR_{k-1}$. For any column $n<k$, the identity $[\mathbf{A}_k]_{k,n}=\prod_{i=n+1}^{k}r_i=r_k[\mathbf{A}_k]_{k-1,n}$ implies that this operation zeroes the $(k,n)$ entry. For column $n=k$, since $[\mathbf{A}_k]_{k,k}=1$ and $[\mathbf{A}_k]_{k-1,k}=a_{k-1}$, the $(k,k)$ entry becomes $1-a_{k-1}r_k$. Hence, within $\mathbf{A}_k$, the last row becomes $[\,\mathbf{0}_{1\times(k-1)},\;1-a_{k-1}r_k\,]$. Since elementary row additions preserve the determinant, expanding along the last row gives the recursion
	\begin{equation}
		\det(\mathbf{A}_k)=(1-a_{k-1}r_k)\,\det(\mathbf{A}_{k-1}),
	\end{equation}
	where $\mathbf{A}_{k-1}$ is the leading $(k-1)\times(k-1)$ submatrix of $\mathbf{A}$, which has the same structure \eqref{eq:A_entries} with base case $\det(\mathbf{A}_1)=1$. Unrolling from $k=2$ to $k=M$ yields
	\begin{equation}
		\det(\mathbf{A})=\prod_{m=1}^{M-1}(1-a_m r_{m+1}).
	\end{equation}
	Substituting back gives
	\begin{equation}
		\begin{aligned}
			\det(\mathbf{Q}_0)
			\!=\!\rho^{M/2}
			\!\!\prod_{m=1}^{M}\!\!\sqrt{\beta_m^{\mathrm{B}}}
			\!\prod_{n=1}^{M}\!\!\sqrt{\delta_n}
			\!\prod_{m=1}^{M-1}\!\!\left(\!1\!-\!\sqrt{\frac{\beta_m^{\mathrm{F}}}{\beta_m^{\mathrm{B}}}}\sqrt{1\!-\!\delta_{m+1}}\,\rho\!\right)\!.
		\end{aligned}
	\end{equation}
	Under the conditions in Proposition~\ref{prop:general_rank}, all three factors are nonzero: $\rho^{M/2}\neq 0$ since $|\rho|=e^{-\alpha_g L/(M+1)}>0$; $\prod_m\!\sqrt{\beta_m^{\mathrm{B}}}\prod_n\!\sqrt{\delta_n}\neq 0$ by $\beta_m^{\mathrm{B}}>0$ and $\delta_n>0$; and each term $(1-a_m r_{m+1})\neq 0$ by hypothesis. Hence $\det(\mathbf{Q}_0)\neq 0$, so $\mathbf{Q}_0$ is invertible and $\mathbf{Q}$ is full row rank, or equivalently $\mathbf{Q}^T$ is full column rank. The DoF conclusion then follows by the same rank-factorization argument as in Appendix~\ref{proof_theorem1}.
	
	\section{Proof of Theorem 2}\label{proof_theorem2}
	We consider that the user is located within a specific service region such that the distance $d_n^{\text{FR}}$ satisfies the condition $d_{\min}^{\text{FR}} \le d_n^{\text{FR}} \le d_{\max}^{\text{FR}}$, for all $n$. Thus, we investigate the power expression as
	\begin{equation}\label{eq:P_R_d}
		\bar{P}_R(d)\! =\! \frac{P_T \eta^2}{2 d^2} \exp\!\big(\! \frac{L\alpha_g }{ M+1}\!\big)\! \underbrace{ \sum_{m=1}^{M}\!  \Bigg| \! \sum_{n=1}^m \!\varrho^{m-n+1} + \!\!\sum_{n=m+1}^{M+1}\!\! \varrho^{n-m} \Bigg|^2 }_{\bar{A}_R},
	\end{equation}
	where $\varrho = \frac{\exp(-\frac{L\alpha_g }{ M\!+\!1} )}{ 2^{1/2} }$, and the received power satisfies
	\begin{equation}
		\bar{P}_R(d_{\max}^{\text{FR}}) \le \bar{P}_R \le \bar{P}_R(d_{\min}^{\text{FR}}).
	\end{equation}
	Then, Eq.~\eqref{eq:P_R_d} can be calculated by
	\begin{equation}
		\begin{aligned}
			& \bar{A}_R \\
			&=\!\!   \left(\! \!  \frac{\varrho}{1\!-\!\varrho}\! \right)^2\! \!  \sum_{m=1}^{M}\! \! \Big( 4 \!+\!\varrho^{2m}\! +\! \varrho^{2M\! -\! 2m\! +\! 2}\! -\! 4 \varrho^m \!-\! 4 \varrho^{M\! -\! m\! +\! 1} \! +\! 2\varrho^{M+1} \Big) \\
			&= \! \! \left(\! \!  \frac{\varrho}{1\!-\!\varrho}\!  \right)^2\! \!  \Big[ 4M \! +\!  2 M \varrho^{M+1}\!  -\!  \frac{8 \varrho (1\! -\! \varrho^M)}{1\! -\! \varrho} + \frac{2 \varrho^2 (1\! -\! \varrho^{2M})}{1\! -\! \varrho^2} \Big].
		\end{aligned}
	\end{equation}
	For large $M$, the variable $\varrho \to \sqrt{1/2}$. Based on this, it can be indicated that
	\begin{equation}
		\bar{A}_R = 4M \left(\! \frac{\varrho}{1\!-\!\varrho}\!  \right)^2 + \mathcal{O}(1).
	\end{equation}
	Then, we can obtain
	\begin{equation}
		\begin{aligned}
			& \frac{P_T \eta^2}{2 (d_{\max}^{\text{FR}})^2} \exp\!\big(\! \frac{L\alpha_g }{ M+1}\!\big)\! \Big[  4M \left(\! \frac{\varrho}{1\!-\!\varrho}\!  \right)^2 + \mathcal{O}(1)\Big] \le \bar{P}_R \\
			& \le \frac{P_T \eta^2}{2 (d_{\min}^{\text{FR}})^2} \exp\!\big(\! \frac{L\alpha_g }{ M+1}\!\big)\! \Big[  4M \left(\! \frac{\varrho}{1\!-\!\varrho}\!  \right)^2 + \mathcal{O}(1)\Big].
		\end{aligned}
	\end{equation}
	Since both the lower and upper bounds scale with $\mathcal{O}(P_T M)$, we conclude that the received power $P_R$ follows the scaling law of $\mathcal{O}(P_T M)$.

	\bibliographystyle{IEEEtran}
	\bibliography{reference/mybib}

\end{document}